\documentclass[final,1p,times]{elsarticle}

\usepackage{subcaption}
\usepackage{graphicx} 
\usepackage{epstopdf}

\usepackage{hyperref}
\hypersetup{
	colorlinks=true,
	linkcolor=cyan,
	filecolor=blue,      
	urlcolor=red,
	citecolor=green,
}

\usepackage{algorithm} 
\usepackage{algorithmicx}
\usepackage{algpseudocode}

\usepackage{float}
\usepackage{booktabs}
\usepackage{multirow}

\usepackage{amsmath}

\journal{Information Sciences}

\begin{document}

\begin{frontmatter}

\title{A lightweight blockchain-based access control scheme for integrated edge computing in the internet of things}

\author[a]{Zhang Jie}\ead{zhangjie@user.ynnu.edu.cn}
\author[a,b]{Yuan Lingyun\corref{*}}\ead{blues520@sina.com}
\author[a,c]{Xu Shanshan}\ead{xushanshan@user.ynnu.edu.cn}
\cortext[*]{ Corresponding author.}

\address[a]{School of Information,Yunnan Normal University,Kunming,650500,China}
\address[b]{Key Laboratory of Educational Information for Nationalities,Ministry of Education,Yunnan Normal University,Kunming,650500,China}
\address[c]{GIS Technology Research Center of Resource and Environment in Western China,Ministry of Education,Yunnan Normal University,Kunming,650500,China}

\begin{abstract}

In view of the security issues of the Internet of Things (IoT), considered better combining edge computing and blockchain with the IoT, integrating attribute-based encryption (ABE) and attribute-based access control (ABAC) models with attributes as the entry point, an attribute-based encryption and access control scheme (ABE-ACS) has been proposed. Facing Edge-Iot, which is a heterogeneous network composed of most resource-limited IoT devices and some nodes with higher computing power. For the problems of high resource consumption and difficult deployment of existing blockchain platforms, we design a lightweight blockchain (LBC) with improvement of the proof-of-work consensus. For the access control policies, the threshold tree and LSSS are used for conversion and assignment, stored in the blockchain to protect the privacy of the policy. For device and data, six smart contracts are designed to realize the ABAC and penalty mechanism, with which ABE is outsourced to edge nodes for privacy and integrity. Thus, our scheme realizing Edge-Iot privacy protection, data and device controlled access. The security analysis shows that the proposed scheme is secure and the experimental results show that our LBC has higher throughput and lower resources consumption, the cost of encryption and decryption of our scheme is desirable.

\end{abstract}


\end{frontmatter}


\newtheorem{thm}{Theorem}
\newdefinition{df}{Definition}
\newdefinition{pf}{Proof}
\newproof{pot}{Proof of Theorem \ref{thm2}}



\section{Introduction}

With the rapid development of modern information technology, mankind is moving towards the era of Internet of Everything, of which the Internet of Things (IoT) is the key technology. It is estimated that there will be more than 41 billion IoT devices by 2027, which is far up from about 8 billion in 2019 \cite{IoTreport}. Another prediction \cite{IoTdata} shows that by 2025, IoT devices will generate approximately 90 ZB of data worldwide. Too much data and devices will be exposed and security is still the problem. Access control is the key technology to protect the data and device safety.\cite{AC-key}

Compared with other traditional access control models, Attribute-Based Access Control (ABAC) model \cite{abac} may be the most suitable for IoT.\cite{AC-key} ABAC grants the access control to the resource (called Object) according to the attributes, which are characteristics that define specific aspects of the subject, object, environment conditions, and/or requested actions that are predefined and preassigned by an authority, presented by a target (called Subject).\cite{abac} Therefore it has the ability to provide more flexible, scalable, secure and fine-grained control for the access request by each IoT device.\cite{abac-1}

However, the distributed architecture of the IoT makes ABAC, where access rights granted by one centralized entity are not suitable. In recent years, some studies have applied blockchain technology (BT) \cite{bitcoin} to realize the distributed access control of the IoT. Blockchain is an distributed, trustless, and secure peer-to-peer (P2P) network, storing data (such as transaction information) by the consensus algorithm in blocks and through hash digests to link them sequentially, in which every single node is equal and has the whole data in the blockchain network.\cite{block-info} With the smart contracts (executable codes that reside in the blockchain), distributed and trustworthy access control can be achieved.\cite{block-sc}

Blockchain based access control can achieve decentralized security but the computational overhead involved is unacceptable for resource-constrained IoT devices.\cite{block-defect} At the same time, if the access of a large number of heterogeneous devices, which generates a large amount of data to be processed in a short time and is simultaneously transmitted to the data center (such as cloud) through the network, and network delay will be inevitable.\cite{net-stress} So the edge computing model for computing the massive amounts of data generated by IoT devices (Edge-Iot) came into being. By putting the computing at the proximity of data sources (edge) and accomplishing tasks such as computing offloading, data storage, caching and processing to solve these problems.\cite{edge-c}

At the one hand, by running the blockchain at the IoT edge node, the computing overhead is solved \cite{edge-bt} and the access control is realized at the same time, at the other hand, data security such as integrity and privacy protection are not completely solved. Cryptology may be a good way to provide data protection. Attribute-Based Encryption (ABE) \cite{abe} was a particular formulation in which data providers will provide a predicate $f()$ in which a formula over the set of string $\chi $ called "attributes" to achieve the data encryption, only the user with the suitable $\chi$ can decrypt a ciphertext encrypted with predicate $f()$.
Ciphertext-Policy Attribute-Based Encryption (CP-ABE) scheme \cite{cpabe} was a public-key encryption scheme based on bilinear pairing and access construction. It is considered as the most suitable technology for providing IoT with data confidentiality and fine-grained access control, for it enables the data owners to define flexible access policy \cite{Fan2019}.

Inspired by these studies, we attempt to introduce the blockchain technology and edge computing to address the limitations of IoT access control. 
Firstly, aiming at the Edge-Iot, current common blockchains such as Bitcoin\cite{bitcoin}, Ethereum\cite{Ethereum}, Fabric\cite{Fabric}, the high resource occupation, difficult in deployment and expansion involved are defective for safe and controllable IoT data access and equipment management, we designed a lightweight blockchain called LBC. The section.\ref{LBC-others} illustrates the comparison result of our LBC and other blockchains. By our LBC, we refer \cite{block-sc} to the implementation of the ABAC model based on smart contracts (SC), called SC-ABAC.

Secondly, based on the CP-ABE, we combine it with ABAC to propose ABE-ACS (Attribute-based encryption and access control scheme, ABE-ACS). We abstract the attributes in ABE and ABAC to achieve the combination of the two, expose the converted access and decrypt policy into the chain and the computationally expensive operations involved during encryption \& decryption phases of CP-ABE are outsourced to edge nodes. Based on ABE-ACS, even if the access policy is matched, the access is still encrypted data, and the attributes must be matched before decryption can be used to obtain the data itself. In contrast, access to the device can only be achieved by matching subject attributes and access policies.

Finally, we implement LBC based on node.js and deployed our LBC on a set of distributed physical machines to simulate the real Edge-Iot environment. By the way that edge nodes act as terminal devices, we conduct thorough experiments based on real data to demonstrate that the proposed ABE-ACS can protect the security of data and provide the devices controllable in practice. 

The evaluation results show that for successful access to data and equipment, thought ABE-ACS, LBC achieves a 300-fold increase in throughput compared to another blockchain under the same configuration,and successful matching the access policy and signature verification with 3 times throughput improvement, penalty mechanism after matching failure (100 times throughput improvement). At the same time, through edge computing, in terms of data encryption and decryption, the time costs of different sizes of data under different numbers of attributes is approximately constant, at the millisecond level, which meets the low latency requirements.

Our contributions are summarized as follows:
\begin{enumerate}[\textbf{.}]
\item We propose a lightweight blockchain called LBC, based on smart contracts we called SC-ABAC to provide secure and controlled access for Edge-Iot.
\item We construct an attribute-based encryption and access control scheme called ABE-ACS, where through edge computing and blockchain technology, ABE-ACS realizes controllable IoT devices and secure data access.
\item We extensively perform experiments using actual IoT heterogeneous devices and data on different platforms. We examine and provide insights on the access control.
\end{enumerate}


\section{Related work}
In this section, we present an overview of the existing research on blockchain-based access control for IoT, and introduce related research on IoT data access control and security with integrated edge computing. 

\subsection{Blockchain in IoT for access control}

With the rapid development in IoT technology, it is not only more important but also more urgent to address secure data access and device management. By leveraging the tamper-proof, accessibility feature and decentralized consensus mechanism of blockchain technology, this problem can be solved.
Blockchain technology is first proposed by Satoshi Nakamoto \cite{bitcoin} in 2008 as a decentralized P2P trading platform. Such as Bitcoin \cite{bitcoin}, Ethereum \cite{Ethereum} , Fabric \cite{Fabric} are three of the most famous blockchain platform. In recent years, several studies and works have been made to use BT for access control in IoT.

Ouaddah et al.\cite{fairaccess} for the first time used BT to implement access control for the IoT, proposed a framework called "FairAccess". In "FairAccess", access tokens are regarded as resource access rights, and access control conditions are met by transferring access tokens, the access token or resource policy is expressed in script language and publicly broadcast to the blockchain network to realize verification and transaction confirmation. By using a Raspberry Pi 2 board with camera as a resource to control access, it demonstrates the usefulness in IoT. However, excessive transaction confirmation and waiting time are still a problem.

Novo \cite{block-info} for the first time used smart contract to implement access control of the IoT, by creating a single smart contract to define all the operations allowed policy rules. By using a set of virtual devices to call a smart contract to return access control results under mutual requests. Zhang et al.\cite{block-sc} used three smart contracts to implement access control of the IoT. Different from Nove\cite{block-info}, Zhang\cite{block-sc} use different smart contracts to implement different functions of access control and implementing with real IoT devices. However, neither of them describes the implementation details of access control.

Maesa et al.\cite{Maesa2017} for the first time to implement ABAC through the blockchain, ABAC strategies are created through scripting language and stored in the chain, and decentralization is achieved by the blockchain transaction strategy. On this basis, Maesa et al.\cite{Maesa2019} converted the logic represented by the ABAC strategy written by eXtensible Access Control Markup Language (XACML, a language is currently available for writing ABAC policies) \cite{XACML} into a smart contract to implement the attribute access strategy and deploy it in the blockchain. However, Maesa only gives an implementation plan, and does not implement it in a specific scene.

Zhang et al.\cite{Zhang2020} aimed at device access control and converted the access policy based on device attributes into an access tree structure to achieve collaborative access. By giving the time-consuming of signature authentication, encryption and decryption of terminal equipment and authentication nodes, as well as the computing overhead of the equipment itself, it proves that the collaborative access control scheme is suitable for IoT devices. However, each access to the device causes the target device to store the keys of both parties, which causes a large storage overhead when multiple devices access a certain device.

Liu et al.\cite{Liu2020} proposed an open-source access control system called "Fabric-iot". In "Fabric-iot", resources are uploaded to the cloud and a url is generated. The user requests the url from "Fabric-iot", among them, the smart contract realizes the function of judging, adding and deleting user attributes and access policies. The processing time is obtained by simulating concurrent access to three smart contracts with multithreaded clients. But for "Fabric-iot", its processing time is greatly affected by the network where the concurrent node is located fluctuations.

Sun et al.\cite{Sun2021} aimed at the problem of accessing cross-domain data in the IoT, introduced the policy decision point (PDP) and policy information point (PIP) in ABAC to obtain and judge the compliance of the access request and the attribute, and finally reached a consensus mechanism for execution. The time-consuming judgment and communication are obtained by simulating the access between different domains. However, in the access request process, the judging mechanism for the repeated joining of the same node in the domain is not considered, and there is a problem of illegal data access.

Song et al.\cite{Song2021} proposed an access control framework called ACF, which uses smart contracts to call traditional access control lists to solve the reliable management of IoT nodes in the supply chain, designs a penalty mechanism under frequent calls to control resources effectively use. By using two blockchain technologies to complete the experimental evaluation. However, access control list is weak to the access control ability, unable to realize fine-grained access control.

Different from the above research, Ding et al.\cite{Ding2019} takes attributes more as the key to access control, using elliptic curve cryptography (ECC) to create public and private key pairs for IoT devices, and encrypting their respective attributes into the blockchain. Access is exchanged through a symmetric key algorithm for access policies, and the access authority is confirmed by the attributes disclosed on the chain. But to establish a one-to-one connection inevitably increases the communication cost.

In summary of the related research, the use of smart contracts can make the implementation of access control more flexible, and the use of the ABAC model can achieve finer-grained access. However, it is worth noting that some blockchain platforms still use centralized Certification Authority (CA) to achieve authentication, making the distribution incomplete. Second, most access control mechanisms only return permission or denial, which does not solve illegal access. In addition, most access control is only for the data itself, but it is more important to forget that restricted devices that are safe and controllable are more important.

\subsection{CP-ABE in IoT for data security}
Sensitive data in the network needs to be secured, and ABE \cite{abe} allows the data owner to define an access policy to encrypt the data, and it can be decrypted only if the policy is met. CP-ABE \cite{cpabe} regards the access policy as the key to encryption, and the visitor generates the corresponding private key based on his own attributes to decrypt. By setting a suitable attribute policy, after encrypting once, the attribute set of multiple users can by decrypted successfully if they matches the policy. However, a large number of resource-limited devices in the IoT environment can not realize the expensive operation in CP-ABE, the main idea of related works is to outsource the computing process to the cloud or edge node.

Cui et al. \cite{Cui2018} proposed a proxy-aided ciphertext-policy ABE (PA-CPABE) outsourcing scheme, which divides the original attribute encryption keys in CP-ABE into public conversion keys (ciphertext conversion by the edge) and the private key of the data owner ensures the security of the key. However, the user's decryption needs to request the cloud first, and then send it to the nearby edge node for initial decryption and then hand it over to the user for secondary decryption, which has high latency and high computational consumption.

Zhang et al.\cite{Zhang2018} proposed an outsourcing and attribute update scheme based on fog nodes. The fog node is used to implement data encryption and decryption by downloading the key and part of the user's key from the cloud. The user only needs to verify and execute it on the resource-constrained device to prove that the computational cost is low. However, the attribute update requires re-encryption of the data, which has high computational consumption.

Fan et al.\cite{Fan2019} proposed an efficient and privacy preserving outsourced multiauthority access control scheme (PPO-MACS) to hide and revoke access policies. User attributes are generated by a one-way hash function, the cloud deletes the stored proxy key after decryption, and the fog node implements outsourcing decryption and verification. However, the security of the fog node itself is unknown, and there is a risk of malicious nodes returning incorrect data.

Li et al.\cite{Li2020} aimed at the problem of attribute duplication in the multi-user access of the IoT, based on the Zhang et al.\cite{Zhang2018} and Fan et al. \cite{Fan2019} proposed the users and attributes of the the revocation. It is also cheaper to perform proof calculations on resource-constrained devices.  However, users can request all ciphertexts from the cloud through the fog node, and there may be a risk of original data leakage.

Aiming at the existing CP-ABE scheme that cannot achieve attribute revocation, attribute addition, outsourcing calculation, and authorization concentration, Sarma et al.\cite{Sarma2021} proposed the PAC-FIT scheme. The user holds a key with the same size. Affected by the number of attributes, the size of the encrypted and decrypted message sent by the user is constant, and the time consumed for encryption and decryption is constant. However, it has not been compared with related researches in the same period, and its program performance does not have high credibility. For example, Tu et al.\cite{Tu2021} also realized attribute revocation and outsourcing calculations. But for the problem of centralized authorization, Tu et al.\cite{Tu2021} adopted multi-authority attribute encryption (MA-ABE), that is, different authorization agencies control different attribute sets. But the comparison of computing performance is also not credible.

Most of the current solutions have too many fully trusted entities such as the Key Generation Center (KGC), Central authority (CA), Attribute manager, etc., so there is a single point of failure and exist the crisis of trust.  However, the introduction of trustless blockchain in combination with CP-ABE is mostly used for data privacy protection.

Hsu et al.\cite{Hsu2020} designed a log protection system based on a private signature chain to CP-ABE, sign and verify data, and store user access records in response to log privacy issues generated by IoT devices.  However, the blockchain verifies the signature for each ciphertext generated before storing it in the block, which causes higher calculation and storage costs.

Guan et al.\cite{Guan2021} proposed privacy-preserving blockchain energy trading scheme (PP-BCETS) in response to the privacy problem of energy decentralized transactions. CP-ABE realizes the privacy protection of transactions, and the blockchain realizes the controllability of decentralized transactions. However, decentralized transactions bring higher computational costs and communication delays.

Qin et al.\cite{LBAC} for the calculation and communication costs caused by verifications of the CP-ABE outsourcing result, the user uses its attribute token as input to call the smart contract to calculate and verify the result, return the corresponding address to access. After decrypting the data, designing credibility and access records to ensure  the data reliable access. However, its high latency and computational time are not suitable for large-scale real-time IoT.

Yang et al.\cite{Yang2021} aimed at the secure sharing of smart grid data, through smart contracts in the blockchain to call edge nodes to implement CP-ABE outsourcing calculations, and use consensus algorithms to ensure correct decryption and secure data sharing. It also gives the effective accountability mechanism for the reliability of edge node calculation results. Our work is partly similar to Yang et al.\cite{Yang2021}. The difference is that our method is more IoT-oriented access control, thinking that CP-ABE cannot achieve access control of devices, and the devices in the IoT are equally important, so we design that through SC-ABAC implements data and device access control, and CP-ABE implements data encryption protection.

The comparison between our study and the above related works is given in the Table.\ref{tab:works2compare}, including using BT for IoT access control and ABE for Edge-Iot data protection. It is evaluated from two aspects of data security and device security. Using blockchain for Edge-Iot and introducing access control models and attribute encryption. Compared with the above research, we have a punishment mechanism for illegal access, fine-grained and real IoT device evaluation environment in access control, at the same time, in terms of encryption, we outsource computing, low latency and low resource consumption, so it is the best in these two aspects.

\begin{table*}[htbp]
  \centering
  \caption{Comparison of Our Study and Other Previous Related Works}
\resizebox{\linewidth}{!}{
    \begin{tabular}{lccccccccc}
    \multicolumn{10}{p{78.955em}}{\textbf{Implementation of Access Control for IoT Data and Devices based on Blockchain}} \\
\toprule
    \multicolumn{1}{c}{\multirow{2}[0]{*}{\textbf{Related Studies}}} & \multicolumn{6}{c}{\textbf{Access Control}}   & \multicolumn{2}{c}{\textbf{Data Security}} & \multicolumn{1}{c}{\textbf{Device Security}} \\
          & \multicolumn{1}{p{6.69em}}{Method} & \multicolumn{1}{p{5.565em}}{Punishment Mechanism} & \multicolumn{1}{p{20em}}{ Implementation Mechanism} & \multicolumn{1}{p{5.44em}}{Blockchain} & \multicolumn{1}{c}{Evaluation} & \multicolumn{1}{p{4.69em}}{Fine-granularity} & \multicolumn{1}{p{5em}}{Data Security Access} & \multicolumn{1}{p{5.75em}}{Data Privacy Protection} & \multicolumn{1}{c}{Device Access Control} \\
\midrule
    Ouaddah et al.\cite{fairaccess} & \multicolumn{1}{p{6.69em}}{Access Token} & \multicolumn{1}{c}{No} & \multicolumn{1}{p{20em}}{Judgment mechanism for passing access token} & \multicolumn{1}{p{5.44em}}{Bitcoin} & \multicolumn{1}{c}{Single IoT Device} & \multicolumn{1}{c}{No} & \multicolumn{1}{c}{Yes} & \multicolumn{1}{c}{No} & \multicolumn{1}{c}{Yes} \\
    Novo \cite{block-info}  & \multicolumn{1}{p{6.69em}}{Access Policy} & \multicolumn{1}{c}{No} & \multicolumn{1}{p{20em}}{Single smart contract defines permission operations} & \multicolumn{1}{p{5.44em}}{Ethereum} & \multicolumn{1}{c}{Virtual IoT Devices} & \multicolumn{1}{c}{No} & \multicolumn{1}{c}{Yes} & \multicolumn{1}{c}{No} & \multicolumn{1}{c}{Yes} \\
    Zhang et al.\cite{block-sc} & \multicolumn{1}{p{6.69em}}{Access Policy} & \multicolumn{1}{c}{Yes} & \multicolumn{1}{p{20em}}{Three smart contract implementation policies to add judgment and punishment} & \multicolumn{1}{p{5.44em}}{Ethereum} & \multicolumn{1}{c}{Two IoT Devices} & \multicolumn{1}{c}{No} & \multicolumn{1}{c}{Yes} & \multicolumn{1}{c}{No} & \multicolumn{1}{c}{Yes} \\
    Maesa et al.\cite{Maesa2017} & \multicolumn{1}{p{6.69em}}{ABAC Model} & \multicolumn{1}{c}{No} & \multicolumn{1}{p{20em}}{Script language implements ABAC policies} & \multicolumn{1}{p{5.44em}}{Bitcoin} & \multicolumn{1}{c}{-} & \multicolumn{1}{c}{Yes} & \multicolumn{1}{c}{Yes} & \multicolumn{1}{c}{No} & \multicolumn{1}{c}{-} \\
    Maesa et al.\cite{Maesa2019} & \multicolumn{1}{p{6.69em}}{ABAC Model} & \multicolumn{1}{c}{No} & \multicolumn{1}{p{20em}}{Smart contracts implement ABAC policies} & \multicolumn{1}{p{5.44em}}{Ethereum} & \multicolumn{1}{c}{-} & \multicolumn{1}{c}{Yes} & \multicolumn{1}{c}{Yes} & \multicolumn{1}{c}{No} & \multicolumn{1}{c}{-} \\
    Zhang et al.\cite{Zhang2020} & \multicolumn{1}{p{6.69em}}{Access Tree} & \multicolumn{1}{c}{No} & \multicolumn{1}{p{20em}}{The smart contracts construct the access policies as access trees} & \multicolumn{1}{p{5.44em}}{Fabric} & \multicolumn{1}{c}{Three IoT Devices} & \multicolumn{1}{c}{Yes} & \multicolumn{1}{c}{-} & \multicolumn{1}{c}{No} & \multicolumn{1}{c}{Yes} \\
    Liu et al.\cite{Liu2020} & \multicolumn{1}{p{6.69em}}{ABAC Model} & \multicolumn{1}{c}{No} & \multicolumn{1}{p{20em}}{Smart contracts implement ABAC policies} & \multicolumn{1}{p{5.44em}}{Fabric} & \multicolumn{1}{c}{Virtual IoT Devices} & \multicolumn{1}{c}{Yes} & \multicolumn{1}{c}{Yes} & \multicolumn{1}{c}{No} & \multicolumn{1}{c}{Yes} \\
    Sun et al. \cite{Sun2021} & \multicolumn{1}{p{6.69em}}{ABAC Model} & \multicolumn{1}{c}{No} & \multicolumn{1}{p{20em}}{Smart contracts calls PDP and PIP to implement ABAC} & \multicolumn{1}{p{5.44em}}{Fabric} & \multicolumn{1}{c}{Single IoT Device} & \multicolumn{1}{c}{Yes} & \multicolumn{1}{c}{Yes} & \multicolumn{1}{c}{No} & \multicolumn{1}{c}{No} \\
    Song et al.\cite{Song2021} & \multicolumn{1}{p{6.69em}}{ACL} & \multicolumn{1}{c}{Yes} & \multicolumn{1}{p{20em}}{Smart contracts implement ACL-based access control} & \multicolumn{1}{p{5.44em}}{Ethereum \& Fabric} & \multicolumn{1}{c}{Proof of Concept} & \multicolumn{1}{c}{No} & \multicolumn{1}{c}{Yes} & \multicolumn{1}{c}{No} & \multicolumn{1}{c}{No} \\
    Ding et al.\cite{Ding2019} & \multicolumn{1}{p{6.69em}}{ABAC Model} & \multicolumn{1}{c}{No} & \multicolumn{1}{p{20em}}{Smart contracts obtain the on-chain attributes that comply with the  ABAC policies} & \multicolumn{1}{p{5.44em}}{Fabric} & \multicolumn{1}{c}{Proof of Concept} & \multicolumn{1}{c}{Yes} & \multicolumn{1}{c}{Yes} & \multicolumn{1}{c}{No} & \multicolumn{1}{c}{No} \\
    Ours  & \multicolumn{1}{p{6.69em}}{ABAC Model} & \multicolumn{1}{c}{Yes} & \multicolumn{1}{p{20em}}{Smart contracts calls PEP,PDP,PAP and PIP to implement ABAC} & \multicolumn{1}{p{5.44em}}{Our LBC} & \multicolumn{1}{c}{Three IoT Devices} & \multicolumn{1}{c}{Yes} & \multicolumn{1}{c}{Yes} & \multicolumn{1}{c}{Yes} & \multicolumn{1}{c}{Yes} \\
\bottomrule
    \multicolumn{10}{l}{\textbf{Yes} support; \textbf{No} not support; \textbf{-} not involved.} \\
\\
    \multicolumn{10}{p{78.955em}}{\textbf{Data Security Access and Privacy Protection for Edge-Iot based on Attribute Encryption Outsourcing}} \\
\toprule
    \multicolumn{1}{c}{\multirow{2}[0]{*}{\textbf{Related Studies}}} & \multicolumn{4}{c}{\textbf{Data Protection}} & \multicolumn{2}{c}{\textbf{Consumption}} & \multicolumn{2}{c}{\textbf{Data Security}} & \multicolumn{1}{c}{\textbf{Device Security}} \\
          & \multicolumn{1}{p{6.69em}}{Method} & \multicolumn{1}{p{5.565em}}{Outsource} & \multicolumn{1}{p{20em}}{ Implementation Mechanism} & \multicolumn{1}{p{5.44em}}{Calculate} & \multicolumn{1}{c}{Latency} & \multicolumn{1}{p{4.69em}}{Resource} & \multicolumn{1}{p{5em}}{Data Security Access} & \multicolumn{1}{p{5.75em}}{Data Privacy Protection} & \multicolumn{1}{c}{Device Access Control} \\
\midrule
    Cui et al. \cite{Cui2018} & \multicolumn{1}{p{6.69em}}{PA-CPABE} & \multicolumn{1}{c}{Yes} & \multicolumn{1}{p{20em}}{The decryption private key is generated from the conversion key and the user private key} & \multicolumn{1}{p{5.44em}}{Fog node} & \multicolumn{1}{c}{H} & \multicolumn{1}{c}{H} & \multicolumn{1}{c}{Yes} & \multicolumn{1}{c}{Yes} & \multicolumn{1}{c}{No} \\
    Zhang et al.\cite{Zhang2018} & \multicolumn{1}{p{6.69em}}{CP-ABE} & \multicolumn{1}{c}{Yes} & \multicolumn{1}{p{20em}}{Decrypted by the combination of the cloud key and the user's partial private key} & \multicolumn{1}{p{5.44em}}{Fog node} & \multicolumn{1}{c}{L} & \multicolumn{1}{c}{H} & \multicolumn{1}{c}{Yes} & \multicolumn{1}{c}{No} & \multicolumn{1}{c}{No} \\
    Fan et al. \cite{Fan2019} & \multicolumn{1}{p{6.69em}}{PPO-MACS} & \multicolumn{1}{c}{Yes} & \multicolumn{1}{p{20em}}{Fog node implements data encryption, decryption and verification} & \multicolumn{1}{p{5.44em}}{Fog node} & \multicolumn{1}{c}{M} & \multicolumn{1}{c}{M} & \multicolumn{1}{c}{Yes} & \multicolumn{1}{c}{Yes} & \multicolumn{1}{c}{No} \\
    Li et al.\cite{Li2020} & \multicolumn{1}{p{6.69em}}{CP-ABE} & \multicolumn{1}{c}{Yes} & \multicolumn{1}{p{20em}}{The ciphertext remains unchanged after revoking the user and his attributes} & \multicolumn{1}{p{5.44em}}{Fog node} & \multicolumn{1}{c}{L} & \multicolumn{1}{c}{L} & \multicolumn{1}{c}{No} & \multicolumn{1}{c}{Yes} & \multicolumn{1}{c}{No} \\
    Sarma et al.\cite{Sarma2021} & \multicolumn{1}{p{6.69em}}{PAC-FIT} & \multicolumn{1}{c}{Yes} & \multicolumn{1}{p{20em}}{The user key is unchanged, not affected by attributes} & \multicolumn{1}{p{5.44em}}{Fog node} & \multicolumn{1}{c}{L} & \multicolumn{1}{c}{-} & \multicolumn{1}{c}{No} & \multicolumn{1}{c}{Yes} & \multicolumn{1}{c}{No} \\
    Tu et al.\cite{Tu2021} & \multicolumn{1}{p{6.69em}}{MA-ABE} & \multicolumn{1}{c}{Yes} & \multicolumn{1}{p{20em}}{Different authorization agencies control different attribute sets} & \multicolumn{1}{p{5.44em}}{Fog node} & \multicolumn{1}{c}{L} & \multicolumn{1}{c}{-} & \multicolumn{1}{c}{Yes} & \multicolumn{1}{c}{Yes} & \multicolumn{1}{c}{No} \\
    Hsu et al.\cite{Hsu2020} & \multicolumn{1}{p{6.69em}}{Private Signature Chain} & \multicolumn{1}{c}{Yes} & \multicolumn{1}{p{20em}}{Private signature chain ensures that encrypted data will not be replaced} & \multicolumn{1}{p{5.44em}}{Fabric} & \multicolumn{1}{c}{H} & \multicolumn{1}{c}{H} & \multicolumn{1}{c}{Yes} & \multicolumn{1}{c}{Yes} & \multicolumn{1}{c}{-} \\
    Guan et al.\cite{Guan2021} & \multicolumn{1}{p{6.69em}}{PP-BCETS} & \multicolumn{1}{c}{No} & \multicolumn{1}{p{20em}}{Transaction data encryption for privacy protection} & \multicolumn{1}{p{5.44em}}{Ethereum} & \multicolumn{1}{c}{H} & \multicolumn{1}{c}{H} & \multicolumn{1}{c}{Yes} & \multicolumn{1}{c}{Yes} & \multicolumn{1}{c}{No} \\
    Qin et al.\cite{LBAC} & \multicolumn{1}{p{6.69em}}{CP-ABE} & \multicolumn{1}{c}{No} & \multicolumn{1}{p{20em}}{Encryption, decryption and verification based on smart contracts} & \multicolumn{1}{p{5.44em}}{Fabric} & \multicolumn{1}{c}{H} & \multicolumn{1}{c}{H} & \multicolumn{1}{c}{Yes} & \multicolumn{1}{c}{Yes} & \multicolumn{1}{c}{Yes} \\
    Yang et al.\cite{Yang2021}  & \multicolumn{1}{p{6.69em}}{CP-ABE} & \multicolumn{1}{c}{Yes} & \multicolumn{1}{p{20em}}{Based on the consensus mechanism to ensure data decryption securely} & \multicolumn{1}{p{5.44em}}{Fabric} & \multicolumn{1}{c}{L} & \multicolumn{1}{c}{L} & \multicolumn{1}{c}{Yes} & \multicolumn{1}{c}{No} & \multicolumn{1}{c}{Yes} \\
    Ours  & \multicolumn{1}{p{6.69em}}{ABE-ACS} & \multicolumn{1}{c}{Yes} & \multicolumn{1}{p{20em}}{Edge node implements data encryption and decryption, consensus mechanism to verificate  results } & \multicolumn{1}{p{5.44em}}{Edge node} & \multicolumn{1}{c}{L} & \multicolumn{1}{c}{L} & \multicolumn{1}{c}{Yes} & \multicolumn{1}{c}{Yes} & \multicolumn{1}{c}{Yes} \\
\bottomrule
    \multicolumn{10}{l}{\textbf{Yes} support; \textbf{No} not support;  \textbf{H} high; \textbf{M} medium; \textbf{L} low; \textbf{-} not involved.} \\
    \end{tabular}%
}
  \label{tab:works2compare}%
\end{table*}%

Overall, we believe that the blockchain itself has insufficient computing power and is not suitable for outsourcing calculations. For edge nodes, only outsourcing calculations are involved. You can refer \cite{Fan2019,Li2020,Sarma2021,Tu2021} for updating and revoking attributes. We provide an implementation of a security solution for Edge-Iot based on blockchain, namely the edge outsourced encryption and decryption of data and the combination of the blockchain. We use the blockchain to honestly record the access process, and design smart contracts to achieve secure access to encrypted data, which is different from Qin et al.\cite{LBAC}. We introduce an access control model based on smart contracts to achieve controllable access to data and equipment. Outsourced computing is confirmed by the Proof of Work (PoW) consensus mechanism. On the basis of ensuring decrypted data, we introduce edge computing to achieve low latency and low computing power consumption. Our LBC test results guarantee high throughput at large scale.


\section{Preliminaries and definitions}

In this section, we first give a brief review of background information on bilinear maps. Then we describe the definition of access structure and relevant background on Linear Secret Sharing Schemes (LSSS) in our paper, and finally we describe CP-ABE and ABAC on which our scheme is based.

\begin{table*}[htbp]
  \centering
  \caption{Description of main notations and acronyms used in this paper}
\resizebox{\linewidth}{!}{
    \begin{tabular}{ll}
\toprule
    \textbf{Notations and Acronyms} & \textbf{Description} \\
\midrule
    $\mathbb{G},\mathbb{G}_{T}$ & Two multiplicative cyclic groups. \\
    $g,p$ & A generator and a prime order of $\mathbb{G}$. \\
    $e$   & A bilinear map $e: \mathbb{G} \times \mathbb{G} \rightarrow \mathbb{G}_{T}$. \\
    $\mathbb{Z}_p,rv$ & The integers modulo $p$, a random value $\in \mathbb{Z}_p$. \\
    $A_S=\{a_1,a_2,\cdots,a_n\}$ & A set of attributes. \\
    $\mathbb{P}=\{PA_s1,PA_s2,\cdots,PA_sn\}$ & Attributes of a set of parties. \\
    $\mathbb{A}$ & Access structure.  \\
    $\mathbb{M}(r,c),$ & A matrix with $r$ rows and $c$ columns. \\
    $\rho(i)$ & A function maps the $i'th$ row of $\mathbb{M}$ to an attribute  set. \\
    $\mathcal{T},\mathcal{T}_r,L_x,NL_x$ & Access tree, root node, leaf node $x$, non-leaf node $x$. \\
    $s=\{s_1,s_2,\cdots,s_n\}$ & A set of secret value. \\
    $PK,PK\_A$ & Public key, public key of A. \\
    $MK$  & Master key. \\
    $pk,pk\_A$ & Private key, private key of A. \\
    $m,CT\_m$ & Message, the ciphertext of message. \\
    $L(x), L(0)$ & \begin{tabular}[c]{@{}l@{}}
A random polynomial of degree $t-1$ to generate \\ secret value $L(0)=s$. 
\end{tabular}
\\
    $Sub,Ob,Op,En$ & Subject, object, operation, environment. \\
    $Sub_A,Ob_A,Op_A,En_A$ & \begin{tabular}[c]{@{}l@{}}
The attributes of the subject, object, operation, \\ and environment respectively. 
\end{tabular}
\\
    $S(t),O(t)$ & \begin{tabular}[c]{@{}l@{}}
A set of active access subjects and passive access objects \\ at time $t$. 
\end{tabular}
\\
    LBC   & Our lightweight blockchain. \\
    Edge-Iot & Internet of things architecture integrating edge computing. \\
    SC-ABAC & \begin{tabular}[c]{@{}l@{}}
Implementation of smart contract based on \\ attribute based access control model. 
\end{tabular}
\\
    ABE-ACS & Attribute-based encryption and access control scheme. \\
\bottomrule
    \end{tabular}%
}
  \label{notations}%
\end{table*}%

We define some notations used in this section. The vectors are denoted by bold letters, and $v_i$ denotes the $i'th$ element of the vector $\vec{\textbf{v}}$. A vector is usually treated as a column vector. $\mathbb{M}^T$ is the transpose of the matrix $\mathbb{M}$. The definitions of main notations and acronyms used in Table.\ref{notations}.

\subsection{Bilinear maps}

\begin{df}\label{def-bilinear}
(Bilinear Maps \cite{bilinear-pairing}).
Let $\mathbb{G},\mathbb{G}_{T}$ be two multiplicative cyclic groups of prime order $p$, and $g$ be the generator of $\mathbb{G}$. The bilinear map $e$ is,
$e: \mathbb{G} \times \mathbb{G} \rightarrow \mathbb{G}_{T}$, for all $a,b\in \mathbb{Z}_{p}$:
\begin{enumerate}[(1)]
\item Bilinearity: $\forall u,v \in \mathbb{G},e\left(u^{a},v^{b}\right)=e(u,v)^{ab}$.
\item Computability: $\forall u,v \in \mathbb{G}$,there is an efficient algorithm to compute $e(u,v)$.
\item Non-degeneracy: $\exists u,v \in \mathbb{G},e(u,v)\neq 1$, where 1 is the unit of $\mathbb{G}$
\end{enumerate}
\end{df}

\subsection{Access structure}

\begin{df}\label{def-acc-stru}
(Access structure \cite{acc-stru}).
Let $\mathbb{P}=\{PA_s1,PA_s2,\cdots,PA_sn\}$ be a set of parties. Set a monotone collection $\mathbb{A}\subseteq 2^\mathbb{P}$. If $\forall B,C$ and if $B\in \mathbb{A}, B\subseteq C$, then $C\in \mathbb{A}$. An access structure is a collection $\mathbb{A}$ of non-empty subsets of $\mathbb{P}$, that is, $ \mathbb{A}\subseteq 2^\mathbb{P}\backslash \{\emptyset \}$. The sets in $\mathbb{A}$ are called the authorized sets, and the sets not in $\mathbb{A}$ are called the unauthorized sets.

\end{df}

In this paper, an access structure $\mathbb{A}$ in ABE contains the authorized sets of attributes. Unless otherwise stated, access structure for the rest of this paper means monotone. 


\subsection{Linear secret-sharing schemes}
\begin{df}\label{lsss}
(Linear Secret-Sharing Schemes \cite{lsss}).
Let $\Pi$ be a secret-sharing scheme over a set of parties $\mathbb{P}$ with realizing an access structure $\mathbb{A}$ is called linear (over $\mathbb{Z}_p$), if:
\begin{enumerate}[1]
\item The shares for each party form a vector over $\mathbb{Z}_p$.
\item There exists a matrix $\mathbb{M}(r,c)$ with $r$ rows and $c$ columns, called the share generating matrix for $\Pi$. For $\forall i \in \{1,2,\cdots,l\}$, using the function $\rho(i)$ to get a set of attribute from the $i'th$ row of $\mathbb{M}$. A vector $\vec{\textbf{v}}=(s,rv_2,\cdots,rv_n)^T$ is generated, where $s \in \mathbb{Z}_p$ is the secret to be shared and $rv_2,\cdots,rv_n \in \mathbb{Z}_p$ are randomly chosen, then $\mathbb{M} \cdot \vec{\textbf{v}}$ is the vector of $l$ shares of the secret $s$ according to $\Pi$. The share $(\mathbb{M} \cdot \vec{\textbf{v}})_i$ belongs to party $\rho(i)$.
\end{enumerate}
\end{df}

In this paper, $\Pi$ is described as $(\mathbb{M},\rho)$. Using standard techniques \cite{lsss} we can convert any monotonic boolean formula into an LSSS representation, such as access binary trees. An access tree of the $l'th$ non-leaf node will result in an LSSS matrix of the $l'th$ row.

\subsection{(t,n) Threshold access tree}
Let $\mathcal{T}$ be a tree representing an access structure with root $\mathcal{T}_r$. Each leaf node $L_x$ of $\mathcal{T}$ is detailed by an attribute, each non-leaf node $NL_x$ of $\mathcal{T}$ is a threshold gate $(t,n)$, which AND or OR gate is the special case. Set $n$ as the number of children of this node $x$ and  $\left(1\leq t \leq n \right) $ is its threshold value. For a threshold gate $(t,n)$, if $t = 1$, the threshold gate is an OR gate and if $t = n$, it is an AND gate. Whether an attribute set $\mathbb{P}$ satisfies a monotone access tree is determined as follows. 

\begin{df}\label{def-acc-tree}

(Threshold access tree \cite{threshold-acc-tree}).
For the leaf node $L_x$, if $L_x \in \mathbb{P}$, the leaf node is said to be satisfied. For a threshold gate $(t,n)$, if and only if at least $t$ (out of $n$ ) child nodes are satisfied, the non-leaf node $NL_x$ is satisfied. If and only if the root $\mathcal{T}_r$ of $\mathcal{T}$ is satisfied, $\mathcal{T}$ is said to be satisfied by $\mathbb{P}$.

\end{df}

In this paper, access policy intuitively express in monotone boolean formulas with AND or OR gates and based on random polynomials to construct a (t,n) threshold access tree and then convert into a matrix by LSSS. Based on the Lagrange interpolation polynomial, the original secret value $s$ of $\mathcal{T}_r$ can only be calculated through at least $t$ values (attributes). 

\subsection{CP-ABE and ABAC}

\subsubsection{Ciphertext-Policy Attribute-Based Encryption (CP-ABE)}
The CP-ABE scheme consists of four algorithms:

$\textbf{Setup}(g,p) \ \rightarrow {PK,MK}$.

The setup algorithm will choose a bilinear group $\mathbb{G}$ of prime order $p$ with generator $g$. Next it will choose two random exponents $pk\_\alpha,pk\_\beta \in  \mathbb{Z}_{p}$. The public key and the master key are published as:

\[ PK=e(g,g)^{pk\_\alpha}, MK=\left(g^{pk\_\alpha},pk\_\beta\right)\]

$\textbf{Encrypt}(PK,m,A_S) \ \rightarrow {CT\_m}$.

The encryption algorithm will take the public key $PK$, a message $m$ and attribute set $A_S$ under the access atructure $\mathcal{T}$ as the input and output the encrypted ciphertext $CT\_m$ of $m$. Assign a value to each node by randomly generating a polynomial $L(x)$, and corresponding the attribute to the leaf node $L_x$. Only the corresponding attribute can solve the value of the root (secret value) to decrypt the ciphertext $CT\_m$. An example of this is given in the section.\ref{lsss-example}.

$\textbf{Key Generation}(PK,MK,A_S) \ \rightarrow {pk}$.

The key generation algorithm will take as input a set of attributes $A_S$ and output a private key $pk$ which described by $A_S$. At the first, it will choose a random $rv \in \mathbb{Z}_{p}$, then $pk$ of different nodes have the same part: 

\[ D = g^{pk\_\alpha / pk\_\beta} * g^{rv / pk\_\beta}\]

Next for $\forall A_S\_i \in A_S$, choose different random $rv\_i \in \mathbb{Z}_{p}$ corresponding to it one by one.

$\textbf{Decrypt}(PK, pk, CT\_m)\ \rightarrow {m}$.

The decryption algorithm will take the public key $PK$, the private key $pk$ and the ciphertext $CT\_m$ as the input, if the attribute set corresponding to the private key $pk$ satisfies the access tree corresponding to the ciphertext $CT\_m$, it will decrypt the ciphertext $CT\_m$ and return the message $m$.

\subsubsection{Attribute-Based Access Control (ABAC) model}
The definition of ABAC model is as follows:
\begin{enumerate}[1)]
\item $Sub,Ob,Op,En$ represent four entities respectively: subject, object, operation, and environment.
\item $Sub_A,Ob_A,Op_A,En_A$ represent the attributes of the subject, object, operation, and environment respectively, such as:
\[Sub_A=\{SA_1,SA_2,\cdots,SA_n\}\]
\[Ob_A=\{Ob_1,Ob_2,\cdots,Ob_n\}\]
\[Op_A=\{Op_1,Op_2,\cdots,Op_n\}\]
\[En_A=\{En_1,En_2,\cdots,En_n\}\]
\item $S(t)=\{s_1,s_2,\cdots,s_n\}$ represents a set of subjects that actively initiate access requests in the IoT at time t, $O(t)=\{o_1,o_2,\cdots,o_n\}$ represents a set of objects that can be accessed in the IoT at time t.
\item Generally, a policy is designed that describes what operations may be performed upon those objects, by whom, and in what environment those subjects may perform those operations. The policy can be expressed as an algorithm that returns a boolean value with the attribute of $Sub,Ob,Op,En$ as input parameters.

$\textbf{Policy Execution}(Sub_i \in S(t),Ob_i \in O(t),Op_i,En_i) \ \rightarrow (True | False)$
\end{enumerate}

If the return value from the policy is true, the subject can perform the operation to object in environment. Otherwise the subject cannot.


\section{System framework and model}
In this section, firstly give the introduce about our lightweight blockchain, and briefly introduce the system framework and model of our scheme.

\subsection{our lightweight blockchain}
Our lightweight blockchain (LBC) is a private chain specially designed for Edge-Iot, similar to current blockchains, our LBC is a traditional chain structure designed by node.js, and the data block linked by the hash pointer is generated by the optimized PoW consensus algorithm. The block stores transaction data, the previous block hash, creation role, creation time, data digest, random number $nonce$ and difficulty $nBits$ of achieving consensus, and the block's own hash. Our LBC uses SHA256, Elliptic Curve Digital Signature Algorithm (ECDSA, using the secp256k1 curve) \cite{ECDSA} to implement hash calculations, data digital signatures, this is the same as most blockchains.

\subsection{System framework}
\begin{figure}[htbp]
\centering
\includegraphics[width=0.99\textwidth]{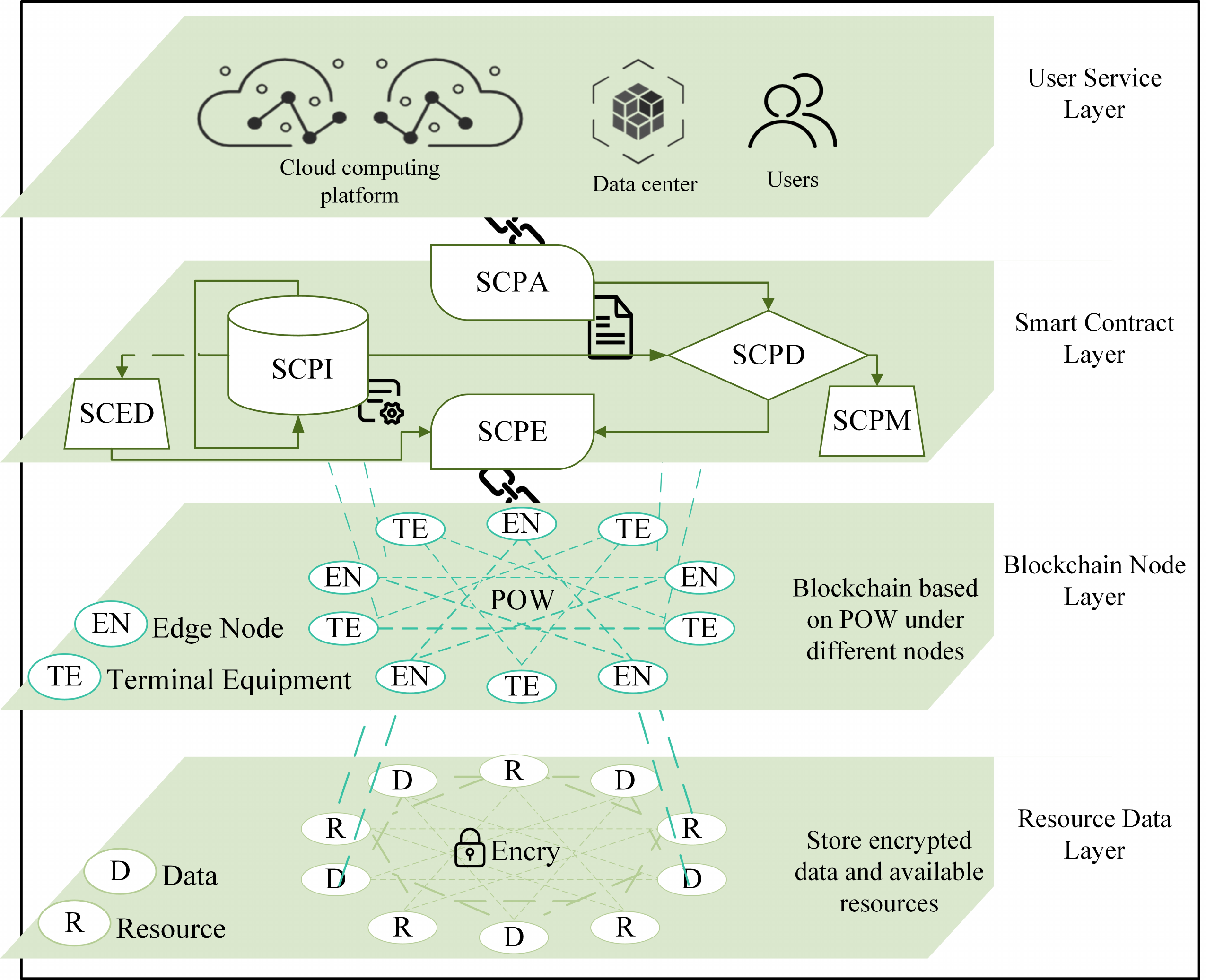}
\caption{The four-layer structural framework followed in our scheme.}
\label{fig:framework}
\end{figure}

We design a four-layer architecture for the realization of our scheme. As is illustrated in Fig.\ref{fig:framework}, the hierarchical structure of the proposed framework can be regarded as the upper two layers of software definition layer and the lower two layers of hardware network layer. The upper layer is implemented by smart contracts to implement access control and penalty mechanisms, and the lower layer is a decentralized blockchain network composed of Edge-Iot, and corresponding resources and data.

\subsubsection{Resource Data Layer}
Resources and data are the objects of access control, which are the resources (such as storage, computing) and data owned by IoT terminal devices and edge nodes that can be provided to other devices or users.

\subsubsection{Blockchain Node Layer}
Blockchain node layer is composed of IoT terminal equipment and edge nodes. Terminal equipments include sensors that can sense environmental data (such as temperature and humidity) and embedded devices that can execute control commands. Edge nodes act as agents for resource-limited devices and receive the data sent by terminal equipment, send the terminal equipment control commands given by the SCPE and realize communication and synchronization through POW consensus and distributed network.

\subsubsection{Smart Contract Layer}
Smart contracts layer is the mechanism guarantee for the implementation of access control, ensuring the management of access policies, the judgment, punishment or execution of object access control, and the management of attributes such as subject and object, etc.
\begin{enumerate}[1)]

\item Smart Contract for Policy Administration (SCPA) :

SCPA is used to manage the access control policies converted from the $(t,n)$ threshold tree and LSSS into the matrix, including adding, modifying, deleting and updating, and to provide encryption policies for data owners. It can only be executed by the policy managers (e.g. the data owners) and sent the policy to SCPD for judgment. An access control policy contains four elements: $Sub_A,Ob_A,Op_A,En_A$. These attributes of elements get by the SCPI.
\item Smart Contract for Policy Decision (SCPD) :

SCPD is used to determine whether the subject's access control complies with the object's corresponding policy. It obtains the subject, object, operation and environmental attributes involved in the current policy from SCPI and returns the judgment result (True or False) to SCPE or SCPM.
\item Smart Contract for Policy Information (SCPI):

SCPI is used to manage the attributes of the subject, object, operation, environment and provide attributes required for data decryption.
$Sub_A$ represents the basic identification information of the subject, such as a unique subject number, identity, etc. $Ob_A$ represents the basic identification information of the object, such as the unique object number, type, etc. $Op_A$ represents the basic information of the operation, such as read, write, delete, create, etc. $En_A$ represents the basic information of the environment, such as time, location, etc.
\item Smart Contract for Encryption Decryption (SCED):

SCED is used for attribute-based encryption and decryption of data, implemented by edge nodes.  For example, encrypt and protect the corresponding attribute of the policy document ontology corresponding to a certain object and decrypt in accordance with the attribute of the subject.
\item Smart Contract for Penalty Mechanism (SCPM):

SCPM is oriented to the subject and used to implement punitive measures under illegal access, such as deleting its corresponding access control policy to make it unable to access other objects, etc.
\item Smart Contract for Policy Enforcement (SCPE):

SCPE is oriented to the object and used to implement legal access control mechanisms, then return the corresponding encrypted data or available resources.

\end{enumerate}

\subsubsection{User Service Layer}
It is used to provide services to users, such as cloud computing, data centers, etc.  Access control is achieved by calling SCPA.

\subsection{System model}
We describe the process through a subject with attributes to access the object under the corresponding policy.  The object's access control policy is transformed from the AND-OR relationship to the $(t,n)$ threshold tree structure and then from LSSS to the corresponding matrix, and the subject decrypts the matrix by matching its own attributes and values, through the Lagrange Interpolation Polynomial to obtain the secret value of the root node and compare it with the real value. If they are equal, the access control policy is satisfied and the object is returned.
\subsubsection{Original Access Control Policy $\mathcal{P}$}

Suppose that there is a subject $Sub$ with an attribute $\mathbb{P}_{Sub}$ such as:
\[\mathbb{P}_{Sub} = \{SA_1, SA_2, SA_3\}\]
At the same time, there is a object $Ob$ with an attribute $\mathbb{P}_{Ob}$ such as:
\[\mathbb{P}_{Ob} = \{ObA_1, ObA_2, ObA_3\}\]
For this object $Ob$, we design an access policy $\mathcal{P}$ based on AND-OR attribute relations:
\[\mathcal{P}\ = \ (SA_1 \ OR \ ObA_1) \ AND \ (SA_2 \ OR \ ObA_2) \ AND \ (SA_3 \ OR \ ObA_3)\]

\subsubsection{Construction based on $(t,n)$ access tree $\mathcal{T}$}
\begin{figure}[htbp]
\centering
\includegraphics[width=0.99\textwidth]{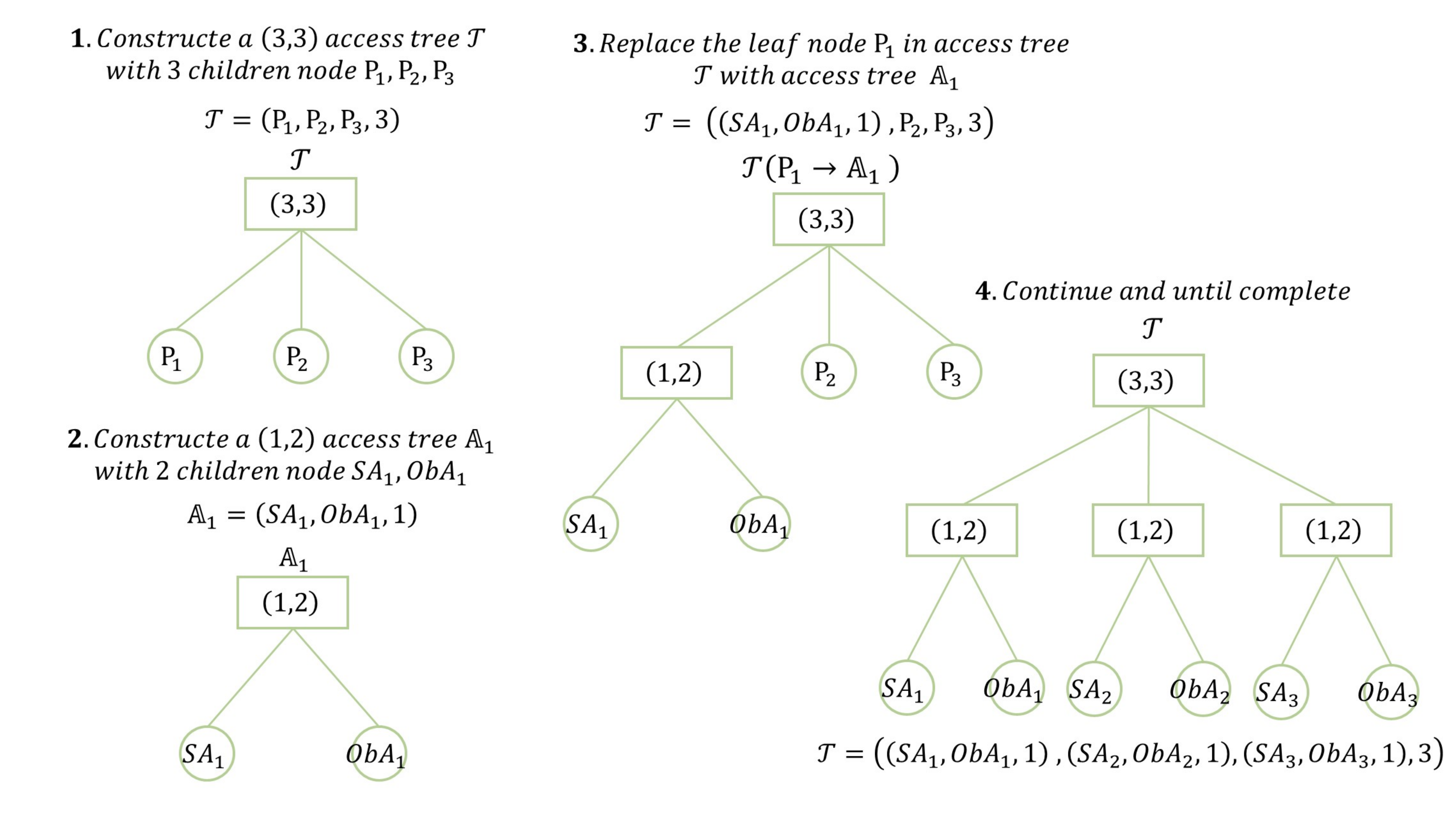}
\caption{Construction of $(t,n)$ access tree $\mathcal{T}$ from access policy $\mathcal{P}\prime$}
\label{(t,n)acctree}
\end{figure}

This access policy $\mathcal{P}$ can be described using a recursive-form string, namely,
\[\mathcal{P}\prime \ = \ ((SA_1,ObA_1,1),(SA_2,ObA_2,1),(SA_3,ObA_3,1),3)\]
Fig.\ref{(t,n)acctree} shows the process from access policy to $(t,n)$ access tree. Finally constructing a (3,3)-threshold access tree with three children node $(SA_1,ObA_1,1)$, $(SA_2,ObA_2,1)$ and $(SA_3,ObA_3,1)$ from $\mathcal{P}\prime$. For the node $(SA_1,ObA_1,1)$, the two children are the leaf nodes $L_x$ corresponding to attributes $SA_1$ and $ObA_1$, which root node is a (1,2)-threshold gate.

\subsubsection{Construction of LSSS-based access structure}\label{lsss-example}

For a $(t,n)$ access tree $\mathcal{T}$, $t$ is its threshold value and $n$ as the number of children of node. Root node $r$ is set with a secret value $\varphi_r$, which the child node is generated by a random polynomial $f(x)$, which the highest power is $t-1$, the secret value $\varphi$ is the constant term, $for \ x = 1,\dots,n$.

When each node $x$ except the root node $r$ calculates its own corresponding secret value $\varphi_{x}$ through the random polynomial $f(x)$, we can construct the corresponding LSSS over $\mathbb{Z}_p$, as

\[
\begin{bmatrix}
t & n & f(1) & f(2) \dots & f(n) \\
t_{1} & n_{1} & f_{1}(1) & f_{1}(2) \dots & f_{1}(n_{1}) \\
t_{2} & n_{2} & f_{2}(1) & f_{2}(2) \dots & f_{2}(n_{2}) \\
\vdots & \vdots & \vdots & \vdots \\
t_{n} & n_{n} & f_{n}(1) & f_{n}(2) \dots & f_{n}(n_{n}) \\
\end{bmatrix}
\ f_{i}(x)=\varphi_{x} \ (for \ i=1,\dots,n; \ x=1,\dots,n_{i})
\]

We now use the access tree shown in Fig.\ref{(t,n)acctree} as an example to construct the LSSS-based structure.
\begin{enumerate}[1)]
\item $\mathcal{T}=\left(7 \right) (\ random \ \varphi_r= 7) \ \leftarrow \mathcal{P}=\left((SA_1,ObA_1,1),(SA_2,ObA_2,1),(SA_3,ObA_3,1),3 \right)$
\item 
$(3,3)-\mathcal{T}=
\begin{pmatrix}
	3 & 3 & 9 & 13 & 19
\end{pmatrix}
 \left(\ random \ polynomial f(x)=x^2+x+7; \ x=1,2,3\right)
 \ \leftarrow \mathcal{P}=
\begin{pmatrix}
	(SA_1,ObA_1,1) \\
	(SA_2,ObA_2,1) \\
	(SA_3,ObA_3,1)
\end{pmatrix}
$
\item 
$(1,2)-\mathbb{A}_1=
\begin{pmatrix}
	1 & 2 & 9 & 9
\end{pmatrix}
 \left(\ random \ polynomial f_{1}(x)=f(1); \ x=1,2\right)
 \ \leftarrow \mathcal{P}\prime=\left(SA_1,ObA_1\right)$

\item
 $final \ \mathcal{T}=
\begin{bmatrix}
	3 & 3 & 9 & 13 & 19 \\
	1 & 2 & 9 & 9 & 0 \\
	1 & 2 & 13 & 13 & 0 \\
	1 & 2 & 19 & 19 & 0 
\end{bmatrix}
$
\end{enumerate}

For calculating the secret value $\varphi_r=7$ in the $(3,3)$ access tree $\mathcal{T}$, suppose that the $Sub$ with an attribute $\mathbb{P}_{Sub} = \{SA_1, SA_2, SA_3\} $, namely $v=
\begin{pmatrix}
	9 & 13 & 19 
\end{pmatrix}
$. 
Then from $f(x)=ax^2+bx+c$ we get a set of equations:
\[
\begin{cases}
a+b+c=9\\
4a+2b+c=13\\
9a+3b+c=19\\
\end{cases}
\]
So we set 
\[
M=
\begin{bmatrix}
	1 & 1 & 1 \\
	4 & 2 & 1 \\
	9 & 3 & 1 \\
\end{bmatrix}
\]
Then compute $M \cdot \omega^T = v^T$, it can be easily computed that $\omega=
\begin{pmatrix}
	1 & 1 & 7 
\end{pmatrix}
$. So the $c=7=\varphi_r$, we can satisfy the access structure. Similarly, using the set of attributes $\mathbb{P}_{Sub}\prime = \{SA_1, SA_2\} $ cannot get the vector $\omega$.


\section{Attribute-based encryption and access control scheme (ABE-ACS)}
\begin{figure}[htbp]
\centering
\includegraphics[width=0.99\textwidth]{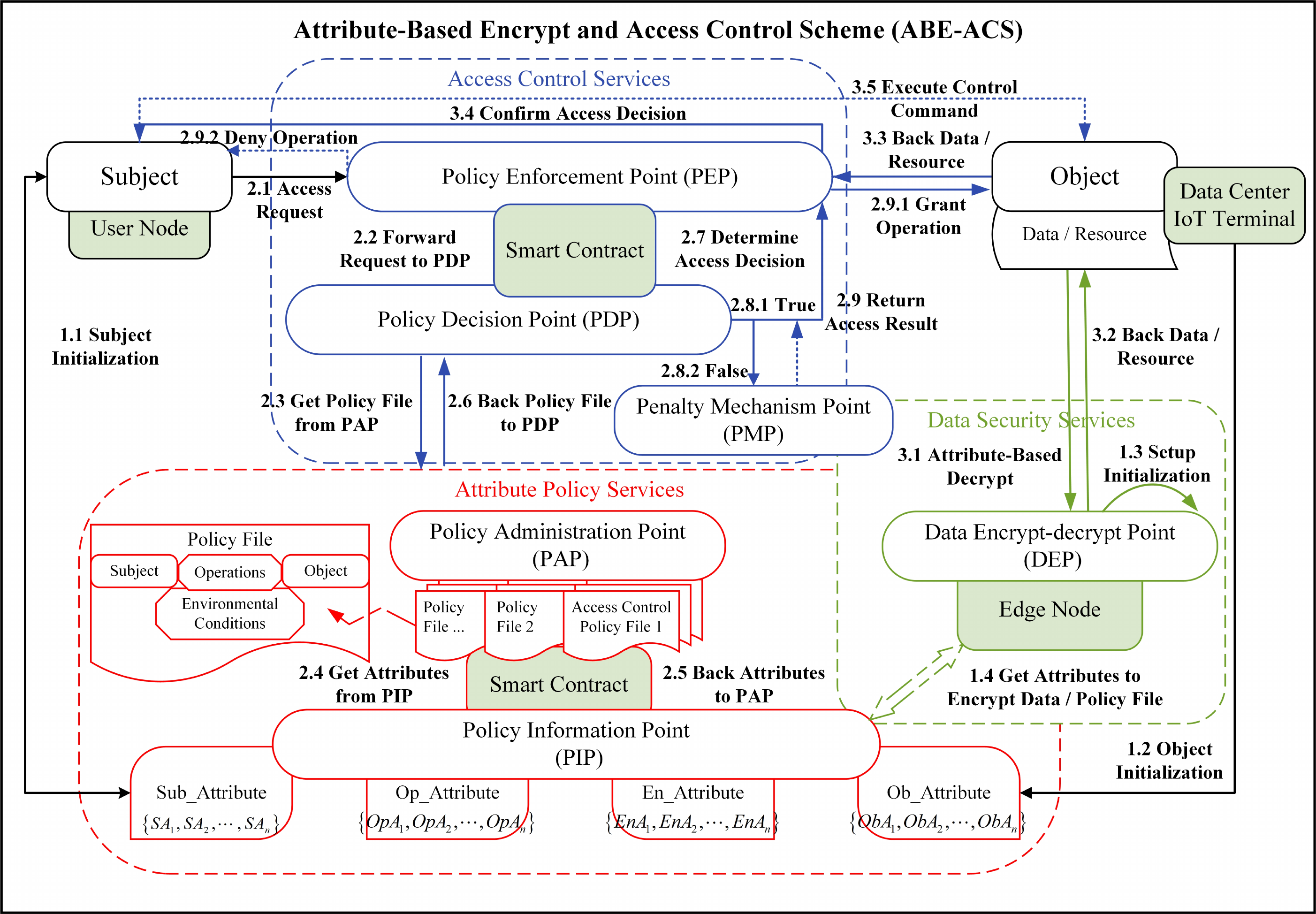}
\caption{Attribute-Based Encrypt and Access Control Scheme (ABE-ACS)}
\label{ABE-ACS}
\end{figure}

In this section, we present the main procedures of the proposed scheme can be divided into the following phases.
\begin{enumerate}[1)]
\item Initialization.
\item Access and control.
\item Outsource.
\end{enumerate}
Our scheme implementation process is shown in Fig.\ref{ABE-ACS}.

\subsection{Initialization}
\subsubsection{Devices and Users Registration}
When a new device wants to apply to join our system, the new node registration is required. Thinking about our scheme for large scale complex networks of heterogeneous devices, which have different ways to communicate, so we identify a device from the four dimensions: device ID:$\mathcal{D}_{ID}$, device COM Port:$\mathcal{D}_{CP}$, device MAC:$\mathcal{D}_{MAC}$ and device IP Port:$\mathcal{D}_{IP}$. We use the SHA-256 to calculate the device ID $\mathcal{D}_{ID}$ and storage it to the labels of nodes in regist set of $\mathcal{D}_{HashID}$. By initializing $rv \in \mathbb{Z}_p$ randomly, when $\mathcal{D}_{ID} \not\in \mathcal{D}_{HashID}$, the $\mathcal{D}_{ID}$ and $\mathcal{D}_{HashID}$ as follows:
\begin{eqnarray}\label{SHA256}
\mathcal{D}_{ID}= SHA_{256}(\mathcal{D}_{ID},\mathcal{D}_{CP},\mathcal{D}_{MAC},\mathcal{D}_{IP},rv)\\
\mathcal{D}_{HashID}=\mathcal{D}_{HashID}.push(\mathcal{D}_{ID})
\end{eqnarray}

After complete execution, a set of new nodes $\mathcal{D}_{ID} \in \mathcal{D}_{HashID}$ is available. Through the unique identification of $\mathcal{D}_{ID}$, the device can be added to the blockchain.
User registration also generates user ID: $\mathcal{U}_{ID}$ and the labels of users in regist set of $\mathcal{U}_{HashID}$. The difference is that the user ID is only determined by the $\mathcal{U}_{ID}$.

\subsubsection{Setup registration}\label{setup}

We set $\mathbb{P}=\{PA_s1,PA_s2,\cdots,PA_sn\}$ as the universe set of attributes, the attributes related to each device and user (such as $Sub_A, Ob_A, Op_A, En_A$) are in it, and they are associated with the corresponding ID. Then set a group $\mathbb{G}=(\mathbb{G}_{1},\mathbb{G}_{T},g,p,e)$ as a bilinear pairing group, let $H:\{0, 1\}^{*} \rightarrow \mathbb{G}$ be a hash function, which maps any attribute to a random element of $\mathbb{G}$. So for a device attribute set $\mathcal{D}A_s=\{DA_1,DA_2,\cdots.DA_n\}$, $H(\mathcal{D}A_s)=\{g^{rv_1},g^{rv_2},\cdots,g^{rv_n}\}$, for randomly value $\{rv_1,rv_2,\cdots,rv_n\} \in \mathbb{Z}_p$. 

We do $H$ mapping to the attribute set in the registered blockchain node, randomly select $g^{pk\_\alpha},g^{pk\_\beta} \in \mathbb{G}$ as the public keys and $pk\_\alpha \in \mathbb{Z}_p$ as the private key for secret preservation.The global public key $PK$ and master key $MK$ we publish to the blocks are set as follows:

\begin{eqnarray}\label{PKMK}
PK=\{\mathbb{G}_{1},\mathbb{G}_{T},g,p,e,g^{pk\_\alpha},g^{pk\_\beta},H\}\\
MK=\{g^{pk\_\alpha},pk\_\beta\}
\end{eqnarray}

\subsection{Access and control}
\subsubsection{Smart Contract for Policy Administration}
\begin{algorithm}[H]
	\caption{Policy Administration.}
	\label{alg:SCPA}
	\begin{algorithmic}[1]
	\Require {User ID:$\mathcal{U}_{ID}$; Labels of users in regist set of $\mathcal{U}_{HashID}$; Device ID:$\mathcal{D}_{ID}$; Labels of nodes in regist set of $\mathcal{D}_{HashID}$}.
	\Ensure {Policy ID:$\mathcal{P}_{ID}$; Labels of policies set of $\mathcal{P}_{HashID}$}.
	\While {$\mathcal{U}_{ID} \subseteq \mathcal{U}_{HashID}$ \&  $\mathcal{D}_{ID} \subseteq \mathcal{D}_{HashID}$}
		\Function {addPolicy}{$\mathcal{U}_{ID},\mathcal{D}_{ID},\cdots$}
			\State  \Return Policy $\mathcal{P}$
		\EndFunction
	\State
		\Function {convertPolicy}{$\mathcal{P}$}
			\State \Return Policy matrix $\mathbb{M}$
		\EndFunction
	\State
		\Function {storePolicy}{$\mathbb{M}$}
			\State \Return $\mathcal{P}_{ID}, \mathcal{P}_{HashID}$
		\EndFunction
	\EndWhile
	\end{algorithmic}
\end{algorithm}

The user manages the policy by calling SCPA, such as adding, converting or storing. Algorithm.\ref{alg:SCPA} shows the relevant operations on the policies. 

$addPolicy()$ takes at least two IDs of subject and object as input, calls the corresponding attributes (mapped by $H$) through $SCPI.getAtt()$ and combine them according to the AND-OR relationship to output a policy file $\mathcal{P}$. 
\begin{algorithmic}[1]
		\Function {addPolicy}{$\mathcal{U}_{ID},\mathcal{D}_{ID},\cdots$}
			\State $\mathcal{U}A_s=SCPI.getAtt(\mathcal{U}_{ID})$
			\State $\mathcal{D}A_s=SCPI.getAtt(\mathcal{D}_{ID})$
			\If {Subject is User and Object is Device}
				\State Select $A_s=\{Sub_A \in \mathcal{U}A_s,\{Ob_A,Op_A,En_A\} \subseteq \mathcal{D}A_s\}$
			\Else
				\State Select $A_s=\{\{Sub_A,Ob_A,Op_A,En_A\} \subseteq \mathcal{D}A_s \}$
				\State Or $A_s=\{\{Sub_A,Ob_A,Op_A,En_A\} \subseteq \mathcal{U}A_s\}$
			\EndIf
			\State Generate Policy $\mathcal{P}$ based on the combination of AND-OR relationship under $A_s$
			\State  \Return Policy $\mathcal{P}$
		\EndFunction
\end{algorithmic}

$convertPolicy()$ takes the $\mathcal{P}$ as input, for easy to understand, our default policy $\mathcal{P}$ itself consists of $Sub_A$, $Ob_A$, $Op_A$ and $En_A$, and they are connected by AND, so split based on AND-OR and recursively return to non-leaf nodes of the threshold tree whose root node is $(4,4)$, and attributes are converted to leaf nodes at corresponding positions. By constructing a random polynomial for each non-leaf node to generate the secret value corresponding to the child node, the leaf node corresponding to the final attribute is assigned one by one, and a matrix $\mathbb{M}$ with a hierarchical relationship and a tree $\mathcal{T}$ are output.
\begin{algorithmic}[1]
		\Function {convertPolicy}{$\mathcal{P}$}
		\State \% Threshold Tree Conversion of Policy
			\State Initialize a tree $\mathcal{T}$ with a root node $(4,4)$
			\State Split $\mathcal{P}$ with AND alone as $x$
			\For {each $x$}
				\If {(AND, OR) $\not\in x$}
					\State The child node $L_x$ of $\mathcal{T}$ is $x$
				\Else
					\State Compute the numbe $n$ of attribute $\in x$
					\If {$\exists OR \in x$}
						\State Initialize a tree $\mathbb{A}$ with a root node $(1,n)$
					\Else
						\State Initialize a tree $\mathbb{A}$ with a root node $(n,n)$
					\EndIf
					\State The child node of $\mathbb{A}$ is $\forall$ attributes $\in x$
					\State The subtree of $\mathcal{T}$ is $\mathbb{A}$
				\EndIf
			\EndFor
		\State \% Matrix Generation based on LSSS
			\State Initialize a matrix $\mathbb{M}=[4,4]$
			\State Randomly select a value $\varphi_{s} \in \mathbb{Z}_p$ to replace the root node of $\mathcal{T}$
			\State Randomly select a set of parameters $\{a_1,a_2,a_3\}(over \in \mathbb{Z}_p)$
			\State Generate polynomial $f(x)=	\varphi_{s}+{a_1}{x}+{a_2}{x^2}+{a_3}{x^3}$
			\For {each child node $NL_i$ of root node$(4,4)$}
				\State $s_i$ for the $i'th \ NL_i \gets f(i)$
				\State the $\mathbb{M}(0,i+1) \gets s_i$
			\EndFor
			\For {each parent node $i'th \ NL_i (t,n)$}
			\State The secret value $s_i$ to replace the $i'th \ NL_i (t,n)$
			\State the $\mathbb{M}(i,0) \gets t;\mathbb{M}(i,1) \gets n$
			\State ${NL_j}\prime \gets$ the $j'th$ child node of $i'th \ NL_i$
			\While {${NL_j}\prime \neq Null$}
				\State Randomly select a set of (t-1) parameters $\{a_1,a_2,\cdots,a_{t-1}\}$
				\State Generate polynomial ${f_i}(x)=s_i+{a_1}{x}+{a_2}{x^2}+\cdots+{a_{t-1}}{x^{n-1}}$
				\State ${s_j}$ for the $j'th \ {NL_j}\prime \gets {f_i}(j)$
				\State the $\mathbb{M}(i,j+1) \gets {s_j}$
			\EndWhile
			\EndFor
			\State \Return Policy matrix $\mathbb{M}$, policy tree $\mathcal{T}$.
		\EndFunction
\end{algorithmic}

$storePolicy()$ takes the $\mathbb{M}$ as input, outputs the policy ID $\mathcal{P}_{ID}$ corresponding to $\mathbb{M}$ by using the SHA-256 to calculate the matrix $\mathbb{M}$ and the secret value $\varphi_{s}$ .
\begin{algorithmic}[1]
		\Function {storePolicy}{$\mathbb{M}$}
			\State $\mathcal{P}_{ID} \gets SHA_{256}(\mathbb{M},\varphi_{s})$
			\If {$\mathcal{P}_{ID} \not\in \mathcal{P}_{HashID}$}
				\State $\mathcal{P}_{HashID}.push(\mathcal{P}_{ID})$
			\EndIf
			\State \Return $\mathcal{P}_{ID}, \mathcal{P}_{HashID}$
		\EndFunction
\end{algorithmic}

When a subject gets the $\mathcal{P}_{ID}$ and $\mathbb{M}$ from the $storePolicy()$, we believe that it is unacceptable to use an exhaustive $\varphi_{s}$ method to calculate the same string as $\mathcal{P}_{ID}$ in the time spent. On the contrary, it is more acceptable to use its own attributes to calculate a corresponding $\varphi_{s}$ and determine whether the calculated string is the same as the $\mathcal{P}_{ID}$ in terms of computational cost. Therefore, the secret value corresponding to any policy can be guaranteed not to be calculated exhaustively, but can only be calculated by the attributes involved in the policy.

\subsubsection{Smart Contract for Policy Decision}
\begin{algorithm}[htb]
	\caption{Policy Decision.}
	\label{alg:SCPD}
	\begin{algorithmic}[1]
	\Require {Subject ID:$\mathcal{S}_{ID}$; Object ID:$\mathcal{O}_{ID}$}.
	\Ensure {The decision: True or False}.
	\While {$\exists \ \mathcal{S}_{ID},\mathcal{O}_{ID}$}
		\Function {judgePolicy}{$\mathcal{S}_{ID},\mathcal{O}_{ID}$}
				\State \Return Decision $Ture$ or $False$
		\EndFunction
	\EndWhile
	\end{algorithmic}
\end{algorithm}

When a subject wants to access an object, for SCPD, it needs to obtain the corresponding attribute set from the SCPI, at the same time to get the corresponding access control policy matrix $\mathbb{M}$ and the $\mathcal{P}_{ID}$ from the SCPA. Therefore, SCPD takes the subject ID $\mathcal{S}_{ID}$ and the object ID $\mathcal{O}_{ID}$ as input, the subject is the requester (such as user, equipment, etc.), and the object is the provider (such as equipment), similarly obtains the attributes $\mathcal{S}A_s$ corresponding to the subject through $SCPI.getAtt()$ and the policy matrix $\mathbb{M}$, policy tree $\mathcal{T}$ and policy ID $\mathcal{P}_{ID}$ corresponding to the object through $SCPA.convertPolicy()$ and $SCPA.storePolicy()$. Then based on the matching of $\mathcal{S}A_s$ and leaf nodes of $\mathcal{T}$, gets the distribution of attributes in the $\mathbb{M}$ and the corresponding values $s_{ij}$, calculates the corresponding secret value $s$ based on $s_{ij}$, combines $\mathbb{M}$ to calculate the string and compare with the $\mathcal{P}_{ID}$, if they are the same, return true, otherwise return false.
Algorithm.\ref{alg:SCPD} shows the relevant operation. 
\begin{algorithmic}[1]
		\Function {judgePolicy}{$\mathcal{S}_{ID},\mathcal{O}_{ID}$}
			\State $\mathcal{S}A_s=SCPI.getAtt(\mathcal{S}_{ID})$
			\State $\mathbb{M},\mathcal{T} \gets $SCPA.convertPolicy()
			\State $\mathcal{P}_{ID} \gets $SCPA.storePolicy()

			\ForAll {$NL_i \in \mathcal{T}$}
				\State Select the same attribute set ${S}A_i \subseteq \mathcal{S}A_s$ as the child attribute node of $NL_i$
				\State The value ${s}_{ij}$ correspinding to ${S}A_i \gets \mathbb{M}(i,j+1)$
				\State $t \gets \mathbb{M}(i,0)$, $n \gets \mathbb{M}(i,1)$
				\State Randomly select a set of (t-1) parameters $\{a_1,a_2,\cdots,a_{t-1}\}$
				\State Randomly polynomial ${f_i}(x)=s\prime+{a_1}{x}+{a_2}{x^2}+\cdots+{a_{t-1}}{x^{n-1}}$
				\For {each $j$}
					\State ${s}_{ij}=s\prime+{a_1}{j}+{a_2}{j^2}+\cdots+{a_{t-1}}{j^{n-1}}$
				\EndFor
				\If {the number of ${S}A_i \geq t$}
					\State $s\prime$ can be computed easily
					\State The secret value $s_i$ of $NL_i \gets s\prime$
				\Else
					\State \Return Decision $False$
				\EndIf
			\EndFor
			\State Randomly select a set of 3 parameters $\{a_1,a_2,a_3\}$
			\State Generate polynomial $f(x)=s+{a_1}{x}+{a_2}{x^2}+{a_3}{x^3}$
			\For {each $i$}
				\State $s_i=s+{a_1}{i}+{a_2}{i^2}+{a_3}{i^3}$
			\EndFor
			\If {the number of $s_i \geq 4$}
				\State $s$ can be computed easily
				\State $\mathcal{P}_{ID}\prime \gets SHA_{256}(\mathbb{M},s)$
				\If {$\mathcal{P}_{ID}\prime \equiv \mathcal{P}_{ID}$}
					\State \Return Decision $True$
				\Else
					\State \Return Decision $False$
				\EndIf
			\Else
				\State \Return Decision $False$
			\EndIf			
		\EndFunction
	\end{algorithmic}
\subsubsection{Smart Contract for Policy Information}
The user manages the attributes by calling SCPI, such as adding, deleting or getting. Algorithm.\ref{alg:SCPI} shows the relevant operations on the attributes.

When a user is successfully registered, he can add or select the corresponding attributes from the universe set of attributes $\mathbb{P}$ by inputting his own $\mathcal{U}_{ID}$, and save the $\mathcal{U}_{ID}$ mapped by the hash function $H$ and the coresponding attribute set $\mathcal{U}A_{s}$ to the $\mathcal{A}_{HashID}$ in the form of key-value pairs, and obtain the corresponding attribute set $\mathcal{U}A_{s}$ by searching for the $\mathcal{U}_{ID}$ or deleting it from the $\mathcal{A}_{HashID}$.
\begin{algorithm}[htb]
	\caption{Policy Informatiion.}
	\label{alg:SCPI}
	\begin{algorithmic}[1]
	\Require {User ID: $\mathcal{U}_{ID}$}.
	\Ensure {The attribute set: $\mathcal{U}A_S$}.
	\State Initialize a universe set of attributes $\mathbb{P}$
	\State Initialize a label of users in attribute sets of $\mathcal{A}_{HashID}$
	\While {$\exists \ \mathcal{U}_{ID}$}
		\Function {addAtt}{User ID: $\mathcal{U}_{ID}$}
			\State Add attributes $A_{s}$ into $\mathbb{P}$
			\State Secert user attribute sets $\mathcal{U}A_{s} \in \mathbb{P}$
			\State $\mathcal{U}A_{s} \gets H(\mathcal{U}A_{s})$
			\State Push $(\mathcal{U}_{ID}, \mathcal{U}A_{s})$ in $\mathcal{A}_{HashID}$ by key-value pair
			\State \Return $\mathcal{A}_{HashID}$
		\EndFunction
\State
		\Function {delAtt}{User ID: $\mathcal{U}_{ID}$}
			\If {$\mathcal{U}_{ID}$ in $\mathcal{A}_{HashID}$}
				\State Delete $(\mathcal{U}_{ID}, \mathcal{U}A_{s})$ from $\mathcal{A}_{HashID}$
				\State \Return $\mathcal{A}_{HashID}$
			\EndIf
		\EndFunction
\State
		\Function {getAtt}{User ID: $\mathcal{U}_{ID}$}
			\If {$\mathcal{U}_{ID}$ in $\mathcal{A}_{HashID}$}
				\State \Return $\mathcal{U}A_{s}$
			\EndIf
		\EndFunction
	\EndWhile
	\end{algorithmic}
\end{algorithm}
\subsubsection{Smart Contract for Penalty Mechanism}

When SCPD determines that the result of a subject's access to the object is false, SCPM is called to return illegal records $\mathcal{I}_{rec}$ and a suitable punitive measure for this subject based the number of illegal access, such as setting the access time limitation, the number of access, etc. Algorithm.\ref{alg:SCPM} shows a punitive measure to limit access time based on illegal access history.

\begin{algorithm}[htb]
	\caption{Penalty Mechanism.}
	\label{alg:SCPM}
	\begin{algorithmic}[1]
	\Require {Subject ID:$\mathcal{S}_{ID}$,Decision:$False$}.
	\Ensure {Punitive measure $P_{M}$,illegal records $\mathcal{I}_{rec}$}.
	\State Initialize a universe set of punitive measures $\mathbb{PM}$
	\State Initialize a illegal records $\mathcal{I}_{rec}$
	\State Initialize a number of illegal access $t \gets 0$ 
	\If {$\exists \ \mathcal{S}_{ID} \in \mathcal{I}_{rec}$}
		\State Get the corresponding vaule $t\prime$
		\State $t \gets t\prime+1$
		\State Delete $(\mathcal{S}_{ID}, t\prime)$ from $\mathcal{I}_{rec}$
	\Else
		\State $t \gets 1$
	\EndIf
	\State Push $(\mathcal{S}_{ID},t)$ in $\mathcal{I}_{rec}$ by key-value pair
	\State Add punitive measures $P_{M}$ into $\mathbb{PM}$
	\State Secert a suitable punitive measure $P_{M} \in \mathbb{PM}$ based the $t$-levels
	\State \% Limit the access time for example
	\State Get the now time in the form of hours (dd:hh) $Time$
	\If {$t \leq 10$}
		\State Next access time $N\_Time \gets Time + 2^t$ (hours)
	\Else
		\State Next access time $N\_Time \gets NULL$
	\EndIf
	\State Push $(\mathcal{S}_{ID},N\_Time)$ in $P_{M}$ by key-value pair
	\State \Return  Punitive measure $P_{M}$,illegal records $\mathcal{I}_{rec}$
	\end{algorithmic}
\end{algorithm}
\subsubsection{Smart Contract for Policy Enforcement}

When SCPD determines that the result of a subject's access to the object is true, or SCPM return the punitive measure, SCPE is called to execute these results. Algorithm.\ref{alg:SCPE} shows the relevant operations on the subject or object.

\begin{algorithm}[htb]
	\caption{Policy Enforcement.}
	\label{alg:SCPE}
	\begin{algorithmic}[1]
	\Require {Subject ID:$\mathcal{S}_{ID}$; Object ID:$\mathcal{O}_{ID}$}.
	\Ensure {Punitive measure $P_{M}$}.
	\State Decision $=SCPD.judgePolicy(\mathcal{S}_{ID},\mathcal{O}_{ID})$
	\If {Decision $\equiv False$}
		\State Get the punitive measure $P_{M}$ and illeagel records $\mathcal{I}_{rec}=SCPM(\mathcal{S}_{ID},Decision)$
		\State Get the now time in the form of hours (dd:hh) $Time$
		\State $t\prime \gets \mathcal{I}_{rec}(\mathcal{S}_{ID})$
		\If {$Time \geq P_{M}(\mathcal{S}_{ID}).N\_Time$}
			\State $t = t\prime - 1$
		\Else
			\State $t = t\prime + 1$
		\EndIf
		\State Delete $(\mathcal{S}_{ID}, t\prime)$ from $\mathcal{I}_{rec}$
		\State Push $(\mathcal{S}_{ID},t)$ in $\mathcal{I}_{rec}$ by key-value pair
		\State \Return Illegal records $\mathcal{I}_{rec}$
	\Else
	\State Get the permission to access the object $\mathcal{O}_{ID}$
	\State \Return The resource of $\mathcal{O}_{ID}$
	\EndIf
	\end{algorithmic}
\end{algorithm}

\subsection{Outsource}

For a visited device, when a visitor obtains the access permission from the PEP, it means that the visitor can control the related resources corresponding to the accessed device's access policy, such as calling the CPU calculation, etc. For the visitor, if obtains the data through the accessed device, still needs to decrypt the data through calling SCED.
The data owner (such as terminal device) can call SCED to encrypt the data by outsourcing it to the edge node, and the corresponding encryption policy may be inconsistent with the policy of the access control process.

The CP-ABE traditional decryption is to input your own private key $pk$ to decrypt the ciphertext $CT\_m$. Because we use the outsourcing encryption and decryption method, the private key cannot be directly handed over to the edge node for decryption, which has security problems. Most of the current outsource methods such as \cite{Outsourcing,Tu2021,Yang2021} are designed to a transformation key $TK$, and decrypting based on the $TK$, user needs only calculate one exponentiation to obtain the massage $m$. However, there are still some resource-limited devices that do not have the ability to perform exponential calculations, so we map the unique data owner (object) ID $\mathcal{O}_{ID}$ by hash function $H()$ to get $g^h$, and then use $g^h$ in both encryption and private key generation process. During the decryption process, the edge node obtains the key $pk\prime$ containing $g^h$ to realize the normal decryption of the data. By generating a new private key $pk\prime$ after each decryption, it is ensured that the private key will not be leaked each time, and the data can be decrypted normally, thereby ensuring the security of the data and the private key.

$encrypt()$ takes public key $PK$, policy matrix $(\mathbb{M},\rho)$, message $m$ and object ID mapping $H(\mathcal{O}_{ID})$ as input and output the ciphertext $CT\_m$.
The specific operation called by the object here is similar to SCPD, but we will give specific instructions in the form of mathematical formulas.
From the section \ref{setup} we get the public key $PK$ and hash function $H()$, and from the $SCPA.convertPolicy()$ we get a matrix $\mathbb{M}(r,c)$, thus, we descript the access policy with $(\mathbb{M},\rho)$, which function $\rho(i)$ gets the attribute value of the $i'th$ row of $\mathbb{M}$. 

For each $i = 1,\cdots,r$, $(t,n) = \mathbb{M}_{i}(0,1)$, we get the matrix $\mathbb{F}$ as follows:
\[
\begin{bmatrix}
1 & 1 &\dots & 1 \\
1 & 2 &\dots & n \\
\vdots & \vdots & & \vdots\\
1 & 2^{t-1} &\dots & n^{t-1} \\
\end{bmatrix}
\]
and choose a random vector $\vec{\textbf{v}_i}=(s_i,a_1,\cdots,a_{n-1})$ (over $\mathbb{Z}_p$), we have: $\vec{\textbf{v}_i}=\rho(i)\cdot\mathbb{F}^{-1}$, the $s_i$ can be calculated, and finally secret value $s$ can be calculated by $s_i$.
In addition, with the $g^h$, the ciphertext $CT\_m$ is published as :

\begin{eqnarray}\label{eq:enc}
CT\_m=\{m \cdot e(g^{pk_{\alpha}},g^h)^s\}, \ C=\{g^{pk_{\beta}\cdot s}\}\\
(C_1=g^{s_1},M_{a1}= H(\rho(1)),\cdots,C_i=g^{s_i},M_{ai}= H(\rho(i)))
\end{eqnarray}

$generateKey()$ takes the public key $PK$, the master key $MK$, the attribute set $A_S$ of subject as input and output the private key $pk$.
We choose random $rv,rv_1,rv_2,\cdots,rv_r \in \mathbb{Z}_p$, let $J \subseteq \{1,\cdots ,r\}$ be defined as $J=\{j:\rho(j) \in A_S\}$. Then for each $j$, by the attribute set $A_S$, the private key $pk$ as follows:
\begin{eqnarray}\label{eq:keygen}
pk=\{g^{(rv+h)/pk_{\beta}}\}, \  D=\{g^{pk_{\alpha}/pk_{\beta}}\}\\
(D_1=g^{rv_1},A_{a1}=g^r\cdot H(J_1)^{rv_1},\cdots,D_j=g^{rv_j},A_{aj}=g^rv\cdot H(J_j)^{rv_j})
\end{eqnarray}

$decrypt()$ takes the public key $PK$, the private key $pk$ and the ciphertext $CT\_m$ as input and output the result of decryption. 
From the $i'th$ row of $\mathbb{M}$ and the $j$ part of $A_S$, computes as follows:
\begin{eqnarray}\label{eq:rsi}
\frac{e(C_i,A_{aj})}{e(D_j,M_{ai})}=\frac{e(g^{s_i},g^rv\cdot H(J_j)^{rv_j})}{e(g^{rv_j},H(\rho(i)))}
=e(g,g)^{rv\cdot s_{i}}, \ if \ J_j=\rho(i)
\end{eqnarray}

For $e(g,g)^{rv\cdot s_{i}},i=1,2,\cdots,r$, using polynomial interpolation can get $e(g,g)^{rv\cdot s\prime}$.
If $s\prime = s$, the message $m$ can be calculated as follows:
\begin{eqnarray}\label{eq:m}
\frac{CT\_m\cdot e(g,g)^{rv \cdot s\prime}}{e(pk,C)\cdot e(D,C)}=
\frac{\{m \cdot e(g,g)^{(pk_{\alpha}+h)\cdot s}\}\cdot e(g,g)^{rv \cdot s\prime}}{\{e(g,g)^{(rv+h)\cdot s}\}\cdot e(g,g)^{pk_{\alpha} \cdot s}}
=m
\end{eqnarray}

By handing over all access control and related operations in the encryption and decryption phases to the edge node, the visitor who matches the attribute policy can obtain the decrypted data calculated by the edge node only by calling SCPE. In order to ensure the authenticity of the data, similar to Yang et al.\cite{Yang2021}, we use PoW to verify the results. Algorithm.\ref{alg:POW} shows the relevant operations on the edge nodes.

Calculate the current block hash $\mathcal{CB}_{Hash}$ including decrypted data $m$ and random number $nNonce$ through multiple nodes, make it satisfy the $nBits$ bits to be $0$. Different from traditional PoW, even if a node first calculates a random number that satisfys the $nBits$ bits to be $0$, other nodes will not stop the calculation, but will verify after the decrypted data is calculated. If it does not meet the requirements, continue to calculate random number $nNonce$ and publish their own calculation results, until most nodes believe that the result is consistent, the consensus is completed. Therefore, the results can be verified by multiple nodes to ensure accuracy.

\begin{algorithm}[H]
	\caption{PoW Improvement.}
	\label{alg:POW}
	\begin{algorithmic}[1]
	\Require {Blockchain difficulty:$nBits$; Previous block hash:$\mathcal{PB}_{Hash}$; Current time:$Time$;Current PoW result:$\mathcal{R}_{PoW}$}.
	\Ensure {Current PoW result: $\mathcal{R}_{PoW}$:((Current block hash:$\mathcal{CB}_{Hash}$;\  Random value:$nNonce$),Number of support:$n$)}.
	\State Message $m=SCED.decrypt()$
	\State $m_{Hash} \gets SHA_{256}(m); nNonce \gets 0$; random value set $R_S$
	\State $\mathcal{CB}_{Hash} \gets SHA_{256}(m_{Hash},\mathcal{PB}_{Hash},Time,nNonce)$	
	\While {$\mathcal{CB}_{Hash}$ first $nBits$ bits $\neq 0$}
		\If {$\mathcal{CB}_{Hash}$ first 1 bit $\neq 0$}
			\State $nNonce \gets nNonce + 1$
		\Else
			\State $nNonce \gets random()$
			\While {$\exists nNonce \in R_S$}
				\State $nNonce \gets random()$
			\EndWhile
		\EndIf
		\State $R_S.push(nNoce)$
	\EndWhile
	\If {$\exists \ $others result $R \in \mathcal{R}_{PoW}$}
		\State Verify result $\mathcal{VB}_{Hash} \gets SHA_{256}(m_{Hash},\mathcal{PB}_{Hash},Time,R.nNonce)$
		\If {$\mathcal{VB}_{Hash} = R.\mathcal{CB}_{Hash}$}
			\State $R.n \gets R.n + 1$
			\State $\textbf{exist}$
		\Else
			\State Push($(\mathcal{CB}_{Hash},nNonce),1$) in $\mathcal{R}_{PoW}$ by key-value pair
		\EndIf
	\Else
		\State Push($(\mathcal{CB}_{Hash},nNonce),1$) in $\mathcal{R}_{PoW}$ by key-value pair
	\EndIf		
	\State \Return Current PoW result: $\mathcal{R}_{PoW}$
	\end{algorithmic}
\end{algorithm}

\section{Security and performance analysis}
In this section, we conduct a comprehensive analysis of the security of ABE-ACS from the following aspects.

\subsection{Formal Verification Through AVISPA Tool}
We use SPAN \cite{SPAN} + AVISPA (Automated Validation of Internet Security Protocols and Applications, AVISPA) simulation tool \cite{AVISPA} to realize and verify the security of our scheme. AVISPA uses the Dolev-Yao threat model to validate the security protocols described using the language HLPSL (High-Level Protocol Specification Language, HLPSL) \cite{HLPSL}. Two validation conclusion, namely: SAFE and UNSAFE, one of them will be returned. 

In the communication process, the communication between edge nodes is recorded by the tamper-free and traceable blockchain, and executed by smart contracts. Therefore, we believe that the process is safe and controllable.  We are aiming for a resource-limit device that may be disguised by a malicious node to monitor and tamper with the communication data between the edge node and the resource-limit devices. We assume that the malicious node has a full control over the local network and can read, store, block every sent message, and can encrypt or decrypt if it has the key. 

\begin{figure}[htbp]
\centering
\begin{subfigure}{0.45\textwidth}
	\centering
	\caption{The OFMC summary report}
	\label{span:OFMC}
	\includegraphics[width=5cm,height=6cm]{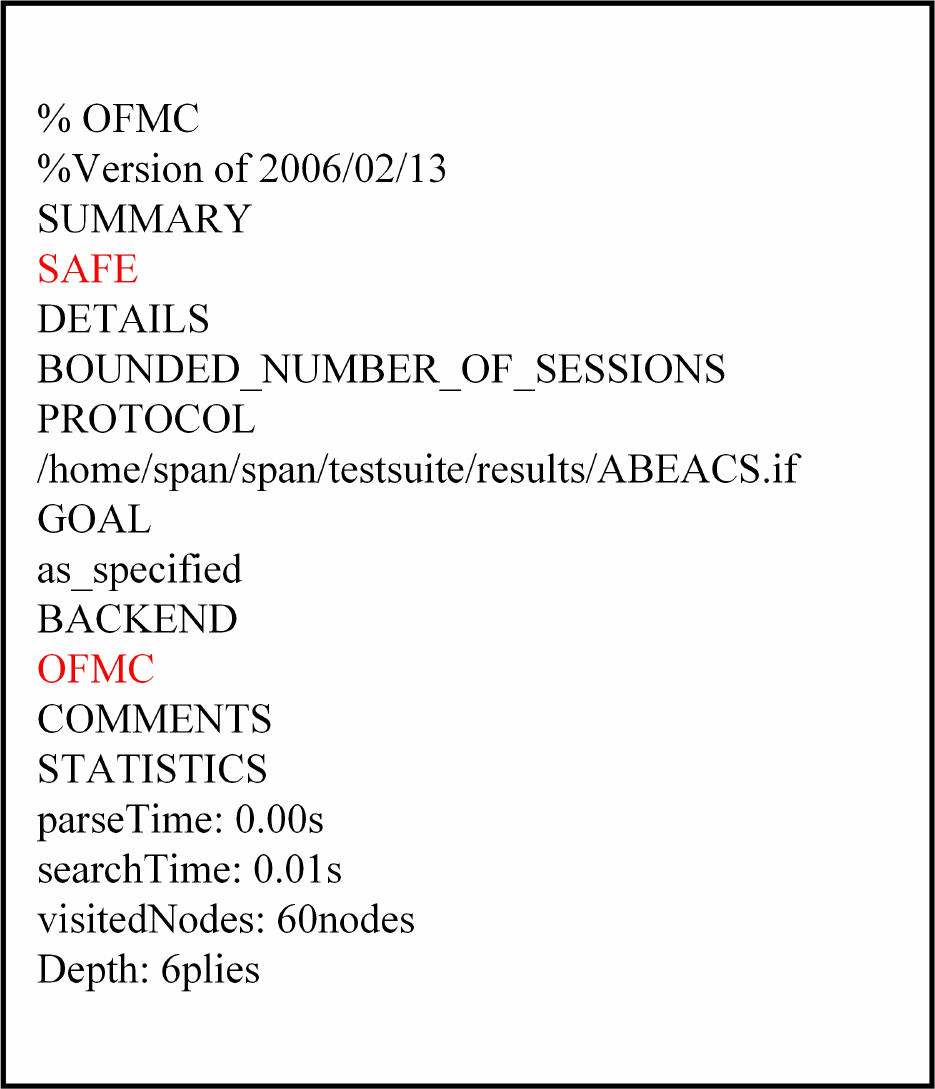}
\end{subfigure}
\begin{subfigure}{0.45\textwidth}{
	\centering
	\caption{CL-AtSe summary report}
	\label{span:CL-AtSe}
	\includegraphics[width=5cm,height=6cm]{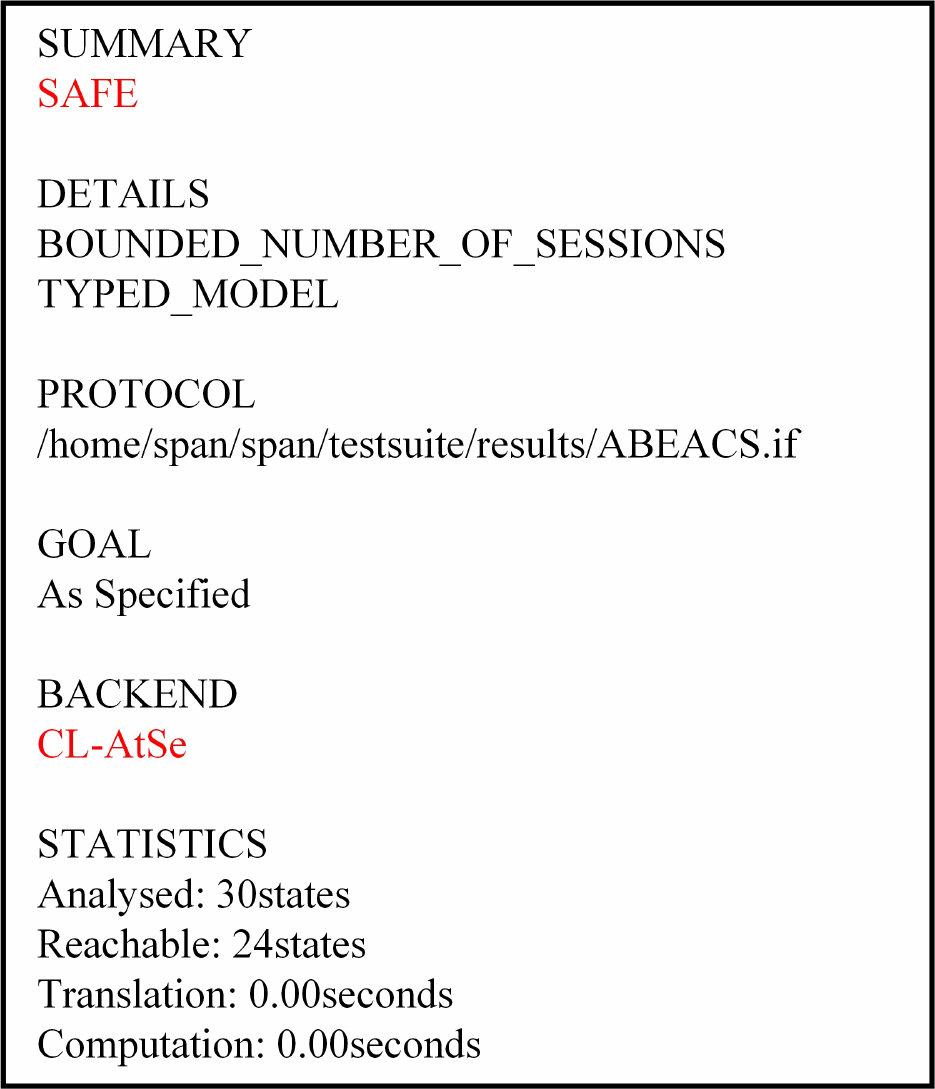}
}
\end{subfigure}
\caption{Analysis of simulation results under OFMC (Fig.\ref{span:OFMC}) and CL-AtSe (Fig.\ref{span:CL-AtSe}) backend.}
\label{fig:span}
\end{figure}

Therefore we use the public-private key pair and random value generated in our scheme to achieve secure communication between the two parties. We used two backends of the AVISPA tool: OFMC and CL-AtSe to validate our scheme. The simulation results of the proposed protocol by using CL-AtSe and OFMC backend of AVISPA tool shows that the proposed scheme is safe as shown in Figure \ref{fig:span} respectively.

\subsection{Theoretical analysis}

\subsubsection{Data secure access}

The heterogeneous environment of the IoT requires higher data security and can ensure the correctness and reliability of the accessed data. Using a distributed blockchain to prevent single points of failure, and to ensure the integrity and immutability of data access records through timestamps and hash indexes. Smart contracts are used to achieve attribute-based fine-grained access control and illegal access punishment mechanisms under the decentralization. At the same time, the optimized PoW consensus algorithm ensures the consistency of outsourcing decryption results. Note that each edge node has the real encrypted data of all resource-limited devices that it represents. Therefore, we do not consider the problem of malicious edge nodes tampering with the data of their own resource-limited devices. We have no way to confirm the correctness and authenticity of the data obtained because of the nature problems of the resource-limited devices. But we can guarantee the secure access to the data in the edge nodes and the authenticity and correctness of the decrypted data.

\textbf{Data confidentiality:} \ Considering that some resource-limited devices cannot achieve exponential decryption capabilities, in order to ensure the security of device private keys in outsourced computing, a mapping of $H()$ to user $\mathcal{U}_{ID}$ is introduced $g^h$, the edge node only completes the decryption after getting the decryption key $pk=g^{(rv+h)/pk_{\beta}}$, and cannot know the real private key $g^{rv}$ through $pk$. At the same time, to re-select the random number set $\{rv,rv_1,rv_2,\cdots,rv_r \in \mathbb{Z}_p\}$ to generate the new poliy and private key $pk$ after each data decryption, and $g^{rv}$ is generated and the data is encrypted again to ensure data confidentiality.  For outsourcing nodes, after computing the decrypted data, $pk=g^{(rv+h)/pk_{\beta}}$ will be invalidated and deleted, ensuring that $pk$ is different for each decryption, ensure the confidentiality of data.

\textbf{Data authenticity:} \ Considering that malicious nodes may deliberately not decrypt data after obtaining $pk$ in order to reduce the time-consuming calculations, and thus complete the first generation of blocks and return false data, the optimized PoW algorithm returns the consistent results calculated by most nodes as the true result, this guarantees the consistency of the decrypted data and at the same time forces the malicious node to complete the calculation to be able to realize the approval of the generated block. On the assumption that most nodes are honest, our optimized consensus mechanism can ensure that even if there are some malicious nodes (all computing power is not more than $50\%$) during the block generation process, the real data can still be obtained. Compared with the traditional PoW consensus algorithm, our optimization algorithm avoids the window-breaking effect caused by malicious nodes, and at the same time, avoids the user's verification of the generated data. Our consensus mechanism can ensure the fairness of generated block and the authenticity of the data.

\textbf{Collusion resistance:} \ Considering that unauthorized users may obtain access rights through collusion, the data owner sets two mechanisms for each data access. One is to encrypt and decrypt according to the attribute policy, and the other is to judge and execute according to the access policy. To achieve collusion resistance, the core is to use a different random number set $\{rv,rv_1,rv_2,\cdots,rv_r \in \mathbb{Z}_p\}$ for each policy, only when the user's own attribute matches the access and decryption policy, the real data can be returned. After decrypting the data, the original data will be encrypted again by the new generated policy. At the same time, the user has four-dimensional attributes $\{Sub_i, Ob_i ,Op_i,En_i\}$ are mapped by $H()$, so unauthorized users are not clear about which attributes satisfy the policies, thereby ensuring collusion resistance.

\subsubsection{Data privacy protection}

The heterogeneous environment of the IoT requires higher privacy protection. Participants of data access need to realize privacy protection for their own attributes and at the same time realize access and obtain data without understanding the policy. The subject can only make an access request on the basis of knowing the object.

\textbf{Attribute anonymity:} \ Although the attribute set $\mathcal{U}A_S$ is public, for the subject and object, the corresponding attribute set $PA_S$ is acquired by $SCPI$, it returns the point set $H(PA_S)=\{g^{rv_1},\cdots,g^{rv_n}\}$ by $H()$ mapped, other nodes cannot know the original attributes even if they query the attribute set $\mathcal{U}A_S$, because using random index ${rv_1,\cdots,rv_n}$ to attribute mapping. So the curious can’t get the real attribute information of other people except for itself, and guarantee the attribute anonymity.

\textbf{Policy anonymity:} \ The key to access control, that is, the policy file is first converted to $(t,n)$ threshold tree and then converted to matrix form $\mathbb{M}$ to store in $SCPA$. Only the hash value of the policy matrix $\mathcal{P}_{ID}$ is returned, and other nodes through the public policy set $\mathcal{P}_{HashID}$ cannot know the specific policy information. Even if multiple nodes comply with the policy file, only the judgment result is returned through $SCPD$, and nothing is known about the policy file itself. Therefore the policy file can be well protected by privacy.

\subsubsection{Device secure access}
In the traditional IoT architecture, resource-limited devices are easily controlled by malicious nodes due to problems such as weak passwords and no authentication mechanisms. By agenting the resource-limit devices through edge nodes, the resource-limit devices only communicate with the edge nodes and do not touch the external network; the communication between the edge nodes is recorded by the blockchain, realized by smart contracts, and guaranteed by consensus mechanism. During the subject's access to the object, the subject can send an access request only when the object is free at the time $t$ and the object $\mathcal{O}_{ID}$ is known. After the object is visited at time $t$, it cannot continue to be visited. At the same time, the smart contract is used to determine the matching of policies and attributes, and achieve access to the object can only be performed on the basis of the subject's attributes $\mathcal{S}A_s$ meeting the access policy $\mathcal{P}$. At the same time, the consensus mechanism ensures that the subject cannot have high-frequency access in a short period of time.

\textbf{Guest authentication:} \ When the system is initialized, for each device, by descirbed and verified by four dimensions $\mathcal{D}_{ID},\mathcal{D}_{CP},\mathcal{D}_{MAC},\mathcal{D}_{IP}$, and at least one of the dimensions met can be registered. The generated random number $rv$ can ensure the uniqueness of the node $\mathcal{D}_{ID}$.
To access a object, the subject must get the object $\mathcal{O}_{ID}$ at the time $t$ by $SCPA$ when the object is free. The attacker cannot forge a certain legitimate node requesting access at a certain moment, except for himself.

\textbf{Controllable access:} \ There are three measures to ensure controlled access. The attributes of the subject $\mathcal{S}A_s$ are described by four dimensions: $\{Sub_i \in S(t), Ob_i \in O(t), Op_i, En_i\}$, minimize the correlation of attributes between different nodes. Different devices have different access policies and different data decryption policies, which can minimize the impression of malicious nodes on themselves. At the same time, the penalty mechanism designed by $SCPM$, such as exponentially increasing the limit access time, can also minimize the frequency of malicious nodes generating requests under random access and the probability of successful access.

\section{Experimental evaluation}

In this section, we present the experimental evaluation results on the performance of ABE-ACS.

\subsection{Experimental Settings}

We implement a prototype of LBC using JavaScript to build the blockchain network, and implement the crypto-graphic mechanism using CP-ABE. 
To deploy blockchain on real IoT environment, we use three computers and three CC2530s. The configuration is shown in Table.\ref{tab:device}.
All tests use the real data collected by data owner with 8051 MCU and 8KB RAM below are conducted on the edge node with 2.90GHz Intel Core i7 processor and 16GB RAM and data manager with 3.40GHz Intel Core i7 processor and 4GB RAM.

\begin{table}[htbp]
  \centering
  \caption{Configuration of experiment}
    \begin{tabular}{ccccc}
    \toprule
    Role  & Type  & CPU   & RAM   & OA \\
    \toprule
    Data owners & CC2530 & 8051 MCU & 8KB   & - \\
    Edge node & Desktop Computer & Intel core i7 (2.90GHz) & 16GB  & Ubuntu 20.04 \\
    Data manager & Server & Intel core i7 (3.40GHz) & 4GB   & DiskStation DS3617xs \\
    \bottomrule
    \end{tabular}%
  \label{tab:device}%
\end{table}%

\begin{figure}[htbp]
\centering
\begin{subfigure}{0.45\textwidth}
	\centering
	\caption{Terminals collect data}
	\label{collectdata}
	\includegraphics[width=6cm,height=5cm]{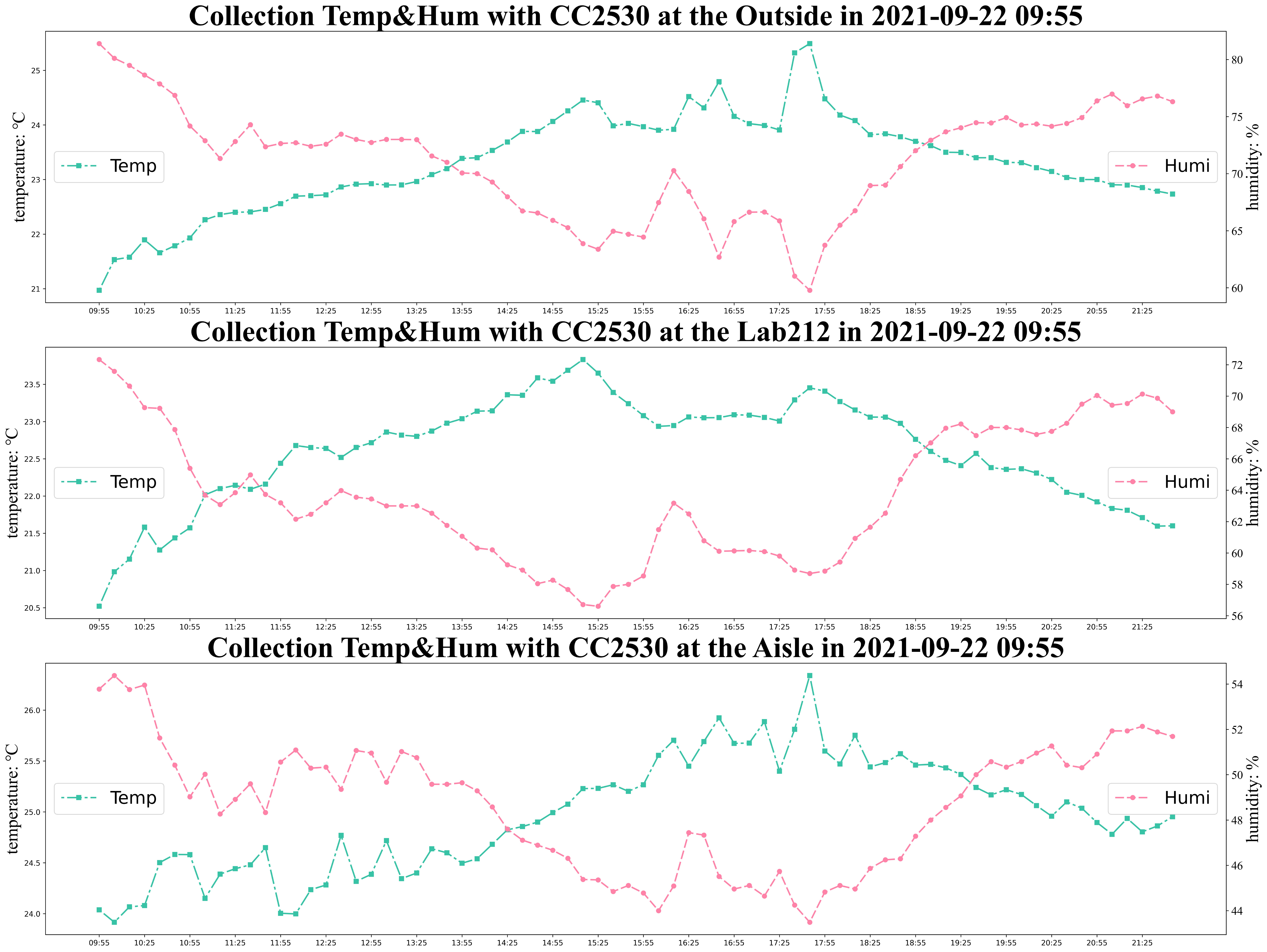}
\end{subfigure}
\begin{subfigure}{0.45\textwidth}{
	\centering
	\caption{Edge node prediction data (Outside)}
	\label{predictedata}
	\includegraphics[width=6cm,height=5cm]{fig7-eps-converted-to.pdf}
}
\end{subfigure}
\caption{The terminals collect temperature and humidity data of three places (Outside, Lab504, Aisle) (Fig.\ref{collectdata}), and edge node processing and prediction (Fig.\ref{predictedata})}
\label{fig:data}
\end{figure}

For actual IoT heterogeneous devices and data, we set three terminals (CC2530) in different places where are outside, lab and aisle to collect the temperature and humidity all the time. Fig.\ref{collectdata} shows the variations of temperature and humidity at the different places in 2021-09-22 09:55, and Fig.\ref{predictedata} shows a simple prediction result and error analysis of the outside temperature and humidity on September 22 obtained by the edge node through the data of the 7 days before the LSTM (Long short-term memory) training.

\subsection{Performance Comparison}\label{LBC-others}

\begin{table*}[htbp]
  \centering
  \caption{The average time to generate a block}
    \begin{tabular}{lcccc}
    \toprule
    Blockchain & Bitcoin\cite{bitcoin} & Ethereum\cite{Ethereum} & Fabric\cite{Fabric} & Our LBC \\
    \midrule
    Time  & 10min & 12.04s & 10ms  & 0.13ms \\
    \bottomrule
    \end{tabular}%
  \label{tab:block-time}%
\end{table*}%

We tested the average block time of three common blockchains at the edge node, shows in table.\ref{tab:block-time}. Due to Bitcoin and Ethereum as public chains, their average block generation time is in the second level, so we only use fabric to build a private chain to compare the performance with our LBC.

\begin{figure}[htbp]
\centering
\includegraphics[width=\textwidth]{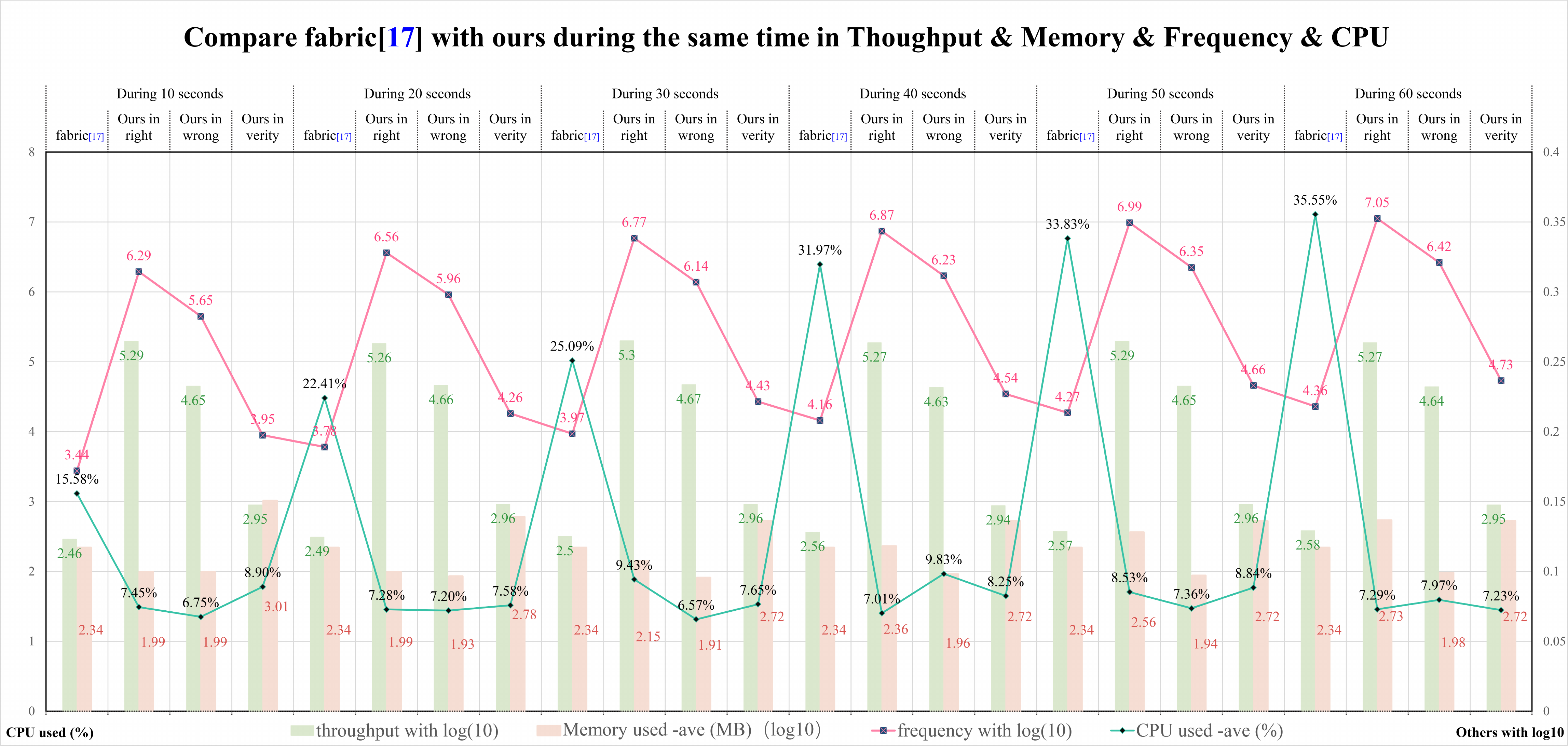}
\caption{Performance comparison of our LBC and fabric \cite{Fabric}}
\label{fig:performance}
\end{figure}

We mainly test transaction performance under duration, such as throughput, memory usage, transaction frequency and CPU usage. Fabric's experimental data comes from Caliper \cite{Caliper}. Our LBC experimental data comes from the sorting of the results of the top command of the edge node linux system. The experimental data is logarithmically standardized except for the CPU usage. Since our LBC is oriented to access control, when comparing the fabric's correct transactions, we conducted three sets of experiments: correct transactions, wrong transactions, and data verification. The test results are shown in the Fig.\ref{fig:performance}.Note that the throughput, storage, and frequency in the figure all use a logarithmic scale.

Fabric has an average throughput of about 300 TPS (transactions per second), and our success is 300 times that of fabric, over 100,000 TPS, and 100 times that of failure, about 45,000 TPS, but the verification process is only 3 times, about 1000 TPS.  This is because we return the result directly when the conditions are met when a successful transaction is met, without verification, and when it fails, there is a penalty mechanism that causes time-consuming. The verification process is to query and verify the signature again, so the verification process has the lowest throughput.  From a time point of view, as the duration increases, the resources occupied by fabric and ours are increasing, but in terms of occupancy, the CPU occupancy of fabric has increased significantly and accounted for a relatively large amount (green line), and ours basically remained stable. However, for the memory, our occupies more memory, which is caused by frequent data exchange in a short period of time.

\subsection{Experimental evaluation}
The main system overheads of our scheme are in two aspects. One is computational costs caused by encryption and decryption, and the another is the time costs caused by generate blocks based on PoW consensus algorithm.
\subsubsection{the CP-ABE }

\begin{table*}[htbp]
  \centering
  \caption{Analysis and comparison of storage and operation costs in devices and edge node}
\resizebox{\linewidth}{!}{
    \begin{tabular}{llllll}
\toprule
          & Role  & Storage overhead & encrypt() & generateKey() & decrypt() \\
\midrule
    \multirow{2}[0]{*}{PAC-FIT\cite{Sarma2021}} & Device(data owner) & $3|\mathbb{G}_{1}|+|\mathbb{G}_{T}|$ & $3\mathbb{G}_{1}+\mathbb{G}_{T}$ & -     & $e$ \\
          & Edge node & -     & $2(1+2N_P)\mathbb{G}_{1}$ & $(4+4N_S)\mathbb{G}_{1}$ & $(2+3N_P)e$ \\
    \multirow{2}[0]{*}{ABE-ACS} & Device(data owner) & $|\mathbb{G}_{1}|$ & $\mathbb{G}_{1}$ & $\mathbb{G}_{1}$ & - \\
          & Edge node & $n\cdot |\mathbb{G}_{1}|$ & $(1+2N_P)\mathbb{G}_{1}+\mathbb{G}_{T}$ & $(2+2N_S)\mathbb{G}_{1}$ & $(2+3N_P)e$ \\
\bottomrule
    \multicolumn{6}{l}{$N_S: $ \ Number of attributes associated with the subject} \\
    \multicolumn{6}{l}{$N_P:$ \  Number of attributes associated with the policy} \\
    \multicolumn{6}{l}{$n:$ \  Number of devices represented by edge node} \\
    \multicolumn{6}{l}{$|\mathbb{G}_{1}|:$ \ Size of the element in $\mathbb{G}_{1}$} \\
    \multicolumn{6}{l}{$|\mathbb{G}_{T}|:$ \ Size of the element in $\mathbb{G}_{T}$} \\
    \multicolumn{6}{l}{$e:$ \ Time required for a bilinear pairing operation} \\
    \multicolumn{6}{l}{$\mathbb{G}_{1}:$ \ Time required for an exponentiation operation in $\mathbb{G}_{1}$ elements} \\
    \multicolumn{6}{l}{$\mathbb{G}_{T}:$ \ Time required for an exponentiation operation in $\mathbb{G}_{T}$ elements} \\
    \end{tabular}%
}
  \label{tab:cpabe}%
\end{table*}%
In Table.\ref{tab:cpabe}, the storage and computational cost of the presented work has been compared with Sarma et al.\cite{Sarma2021}. In storage overhead, the devices only need to store a $g^{rv}$, so just $|\mathbb{G}_{1}|$ of data, and for edge node, in which represents $n$ devices, during the decryption needs to store the private key $pk=g^{(rv+h)/pk_{\beta}}$ of each device, there are $n\cdot |\mathbb{G}_{1}|$ of data, which is may more than the method of Sarma et al. This is because we try to reduce the storage pressure of the devices. For operation costs, during the $generateKey()$ phase, devices need to calculate the $g^{(rv+h)}$, which is $\mathbb{G}_{1}$. Edge node needs to calculate the values $D_i,A_{ai}$ of each attribute of the policy and the $pk$, $D$, which is $(2+2N_S)\mathbb{G}_{1}$. This is half consumption of Sarma et al. During the $encrypt()$, there is nothing to do for devices that have data, and for subject needs to calculate $g^h$, which is $\mathbb{G}_{1}$. Edge node needs to calculate the value $C_i, M_{ai}$ of each attribute of the subject and the $CT_m$, $C$, which is $(1+2N_P)\mathbb{G}_{1}+\mathbb{G}_{T}$, also nerly half. During the $decrypt()$, diffierent with Sarma et al, there are nothing to do for devices, and edge node is same as the method of Sarma et al in consumption.

\begin{figure}[htbp]
\centering
\begin{subfigure}[encryption]{1\textwidth}{
	\centering
	\caption{Encryption}
	\includegraphics[scale=0.25]{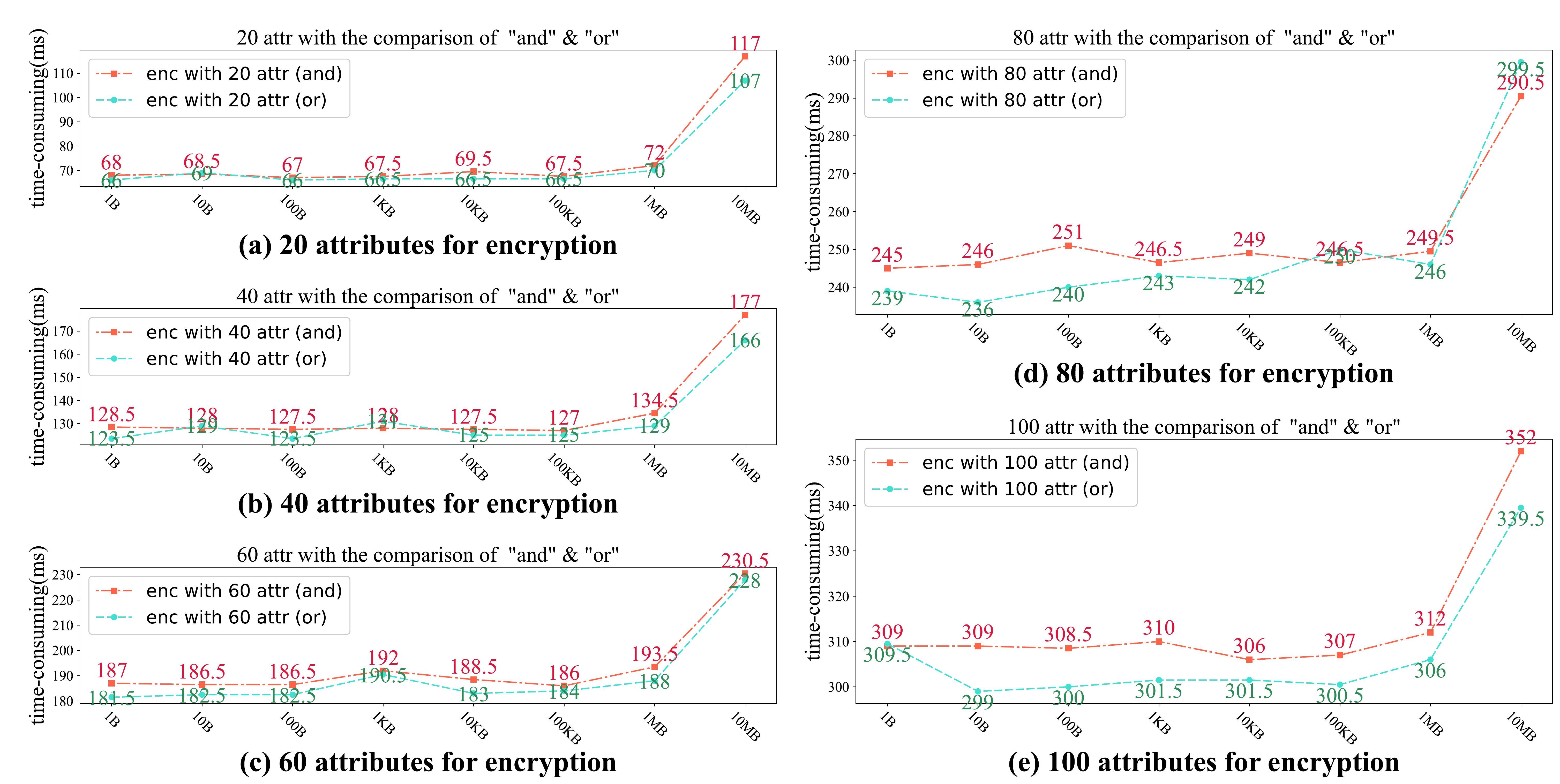}
}
\end{subfigure}
\\
\begin{subfigure}[decryption]{1\textwidth}{
	\centering
	\caption{Decryption}
	\includegraphics[scale=0.25]{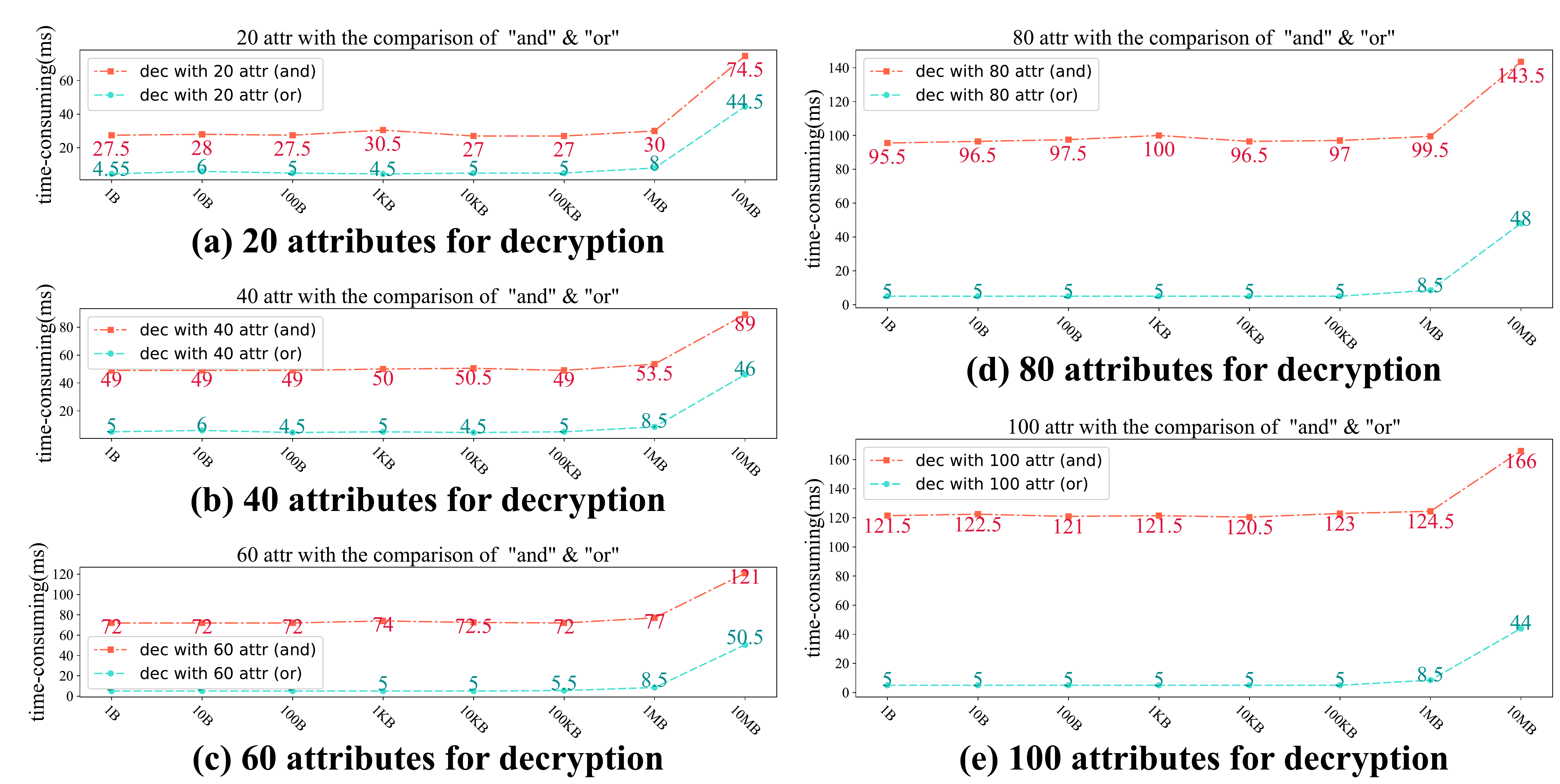}
}
\end{subfigure}
\caption{The final encryption(a) and decryption(b) time-consuming with different size data and the number of attributes}
\label{en-de}
\end{figure}

In performance testing, we test the encryption and decryption time with the CP-ABE at the edge node. We set the data size from 1B to 10MB, show the encryption and decryption time with different number of attributes based and-or with the CP-ABE.
As shown in Fig.\ref{en-de}, we compare the encryption and decryption time-consuming from three aspects: the number of attributes, the data size, the way to encrypt or decrypt. For the number of attributes, with the number of attributes increases, the encryption and decryption time-consuming is linear increasing. For the data size, when the data size is less than 1MB, the time-consuming mostly can be regarded as a constant; for the way to encrypt or decrypt, "and" or "or" in encryption is similar, but in decryption, if we choose "or",the time does not change with the number of attributes, and "and", decryption time-consuming is linear increasing with the number of attributes increases.

\subsubsection{the PoW's algorithms}

\begin{figure}[htbp]
\centering
\includegraphics[width=0.99\textwidth]{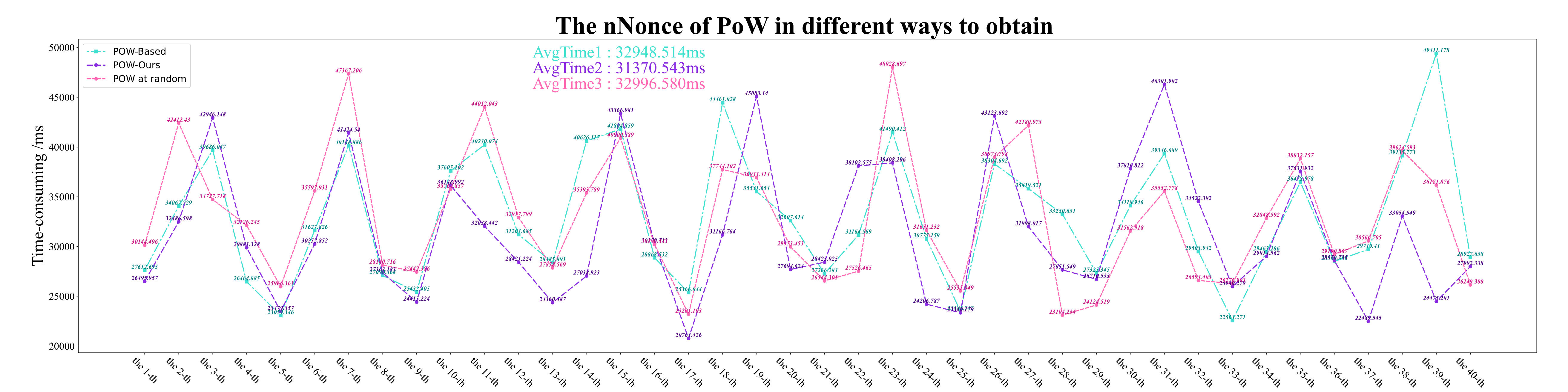}
\caption{The time-consuming about three PoW's algorithm. The comparison among PoW-Based, PoW-randomly and PoW-Ours with the $nNonce=5$.}
\label{PoW}
\end{figure}

For the three PoW's algorithm, in the case of  $nNonce = 5$, on the data manager node we set up four sets of controlled experiments, namely single, three times of concurrency, four times of concurrency and five times of concurrency. Each group is performed 100 times to find the average time-consuming, and 10 tests totally. And showing the order in the figure is once, four-time, five-time and three-time, time-consuming shows mostly the five-time concurrency is the highest, and the once is the lowest in the four groups. At the same time, the average block-produced time-consuming does not much different, PoW-Ours about 31s per block, and PoW-Based about 33s per block. However, from the Fig.\ref{PoW}, PoW-Ours from the once is 21s per block to the five-time concurrency is 45s per block, so if we make the peers choose the low concurrency, PoW-Ours is the best.

\section{Conclusion}

This paper proposes an attribute-based encryption and access control scheme (ABE-ACS). Under the security issues of Edge-Iot, our scheme implements ditributed and trustworthy  access control based on blockchain and performs outsourcing decryption based on edge computing. 
Our scheme is executed by multiple smart contracts to realize access control and penalty. Improved PoW consensus mechanism ensures the correctness of outsourcing decryption without additional computing and communication costs on the user side. 
Moreover, we design a lightweight blockchain (LBC) for the implementation of smart contracts. The combination of the $(t,n)$ threshold tree and LSSS enables privacy protection of the policy. A new CP-ABE scheme ensures private key protection for resource-limited devices.
We gave the security analysis to prove the security of our scheme. In the performance evaluation, we tested from three aspects and compared our scheme with related schemes, the results show that our scheme is efficient and practical.
In the future, the focus of our work is to realize the automatic management of attributes and the automatic generation of policies.





\section*{CRediT author statement}
\textbf{Zhang Jie:} Conceptualization, Methodology, Formal analysis, Writing- Original, Visualization. \textbf{Yuan Lingyun:} Conceptualization, Resources, Writing - Review \& Editing, Project administration, Funding acquisition. \textbf{Xu Shanshan:} Software, Validation, Investigation, Data Curation.

\section*{Declaration of Competing Interest}
The authors declare that they have no known competing financial interests or personal relationships that could have appeared to influence the work reported in this paper.

\section*{Acknowledgment}
This work is partially supported by the National Natural Science Foundation of China (No.61561055), the Special Fundamental Research Project of Yunnan Province-General Project (No.202101AT070098), The Yunnan Ten-thousand Talents Program, and the Graduate Innovation Fund of Yunnan Normal University (No.ysdyjs2020148).


\end{document}